\documentclass[onecolumn,authoryear]{els-mrw}

\usepackage{amsmath,amssymb,amsfonts,amsthm,makeidx,graphicx}
\usepackage{txfonts}
\usepackage{helvet}
\usepackage[colorlinks=true,citecolor=blue, urlcolor=blue, ]{hyperref}
\usepackage[normalem]{ulem}
\usepackage{xcolor}

\DeclareRobustCommand{\Eqref}[1]{Equation~\ref{#1}}
\DeclareRobustCommand{\Figref}[1]{Figure~\ref{#1}}

\DeclareRobustCommand{\Secref}[1]{Section~\ref{#1}}

\newcommand{\cag}{$^{12}\mathrm{C}(\alpha,\gamma)^{16}\mathrm{O}$~}
\usepackage{soul}

\def\aj{AJ}
\def\apj{ApJ}
\def\apss{Astrophys.~Space Sci.}
\def\apjl{ApJL}
\def\apjs{ApJS}

\def\aap{A\&A}
\def\araa{ARAA}
\def\mnras{MNRAS}

\def\prl{PRL}
\def\prc{PRC}
\def\prd{PRD}

\def\pasp{PASP}

\def\nat{Nature}

\def\zap{Zeit.~Astr.}
\def\pasj{Pub.~Astr.~Soc.~Japan}

\begin{document}

\graphicspath{{./figures/}}

\chapter{Pair-instability evolution and explosions in massive stars}\label{ch:pair-instability}

\author[1]{M.~Renzo}
\author[1]{N.~Smith}

\address[1]{\orgname{University of Arizona}, \orgdiv{Department of
    Astronomy \& Steward Observatory}, \orgaddress{933 N. Cherry Ave., Tucson, AZ 85721, USA}}

\articletag{(P)PISN}

\maketitle

\begin{glossary}[Glossary]
  \term{pair} electron ($e^{-}$) and positron ($e^{+}$) pair, also $e^{\pm}$\\
  \term{supernova} luminous electromagnetic transient\\
  \term{SN~Ia} supernova with spectrum showing no hydrogen lines (from the explosion) but showing silicon lines\\
  \term{SN~Ib} supernova with spectrum showing helium but not hydrogen lines\\
  \term{SN~Ic} supernova with spectrum showing neither helium nor hydrogen lines\\
  \term{SN~II} supernova with spectrum showing hydrogen lines (and no silicon lines)\\
  \term{SN~IIn} SN~II with narrow emission lines\\
  \term{SN~Ibn} SN~Ib with narrow emission lines\\
  \term{metallicity} fraction of mass in elements heavier than hydrogen and helium\\
\end{glossary}

\begin{glossary}[Nomenclature]
  \begin{tabular}{@{}lp{34pc}@{}}
    SN & Supernova\\
    SLSN & Superluminous supernova (peak absolute magnitude $<21$)\\
    PISN & Pair instability supernova\\
    P-PISN & Pulsational pair instability supernova\\
    BH & Black hole\\
    GW  & Gravitational wave\\
    Z & Metallicity, that is mass fraction of elements heavier than helium\\
    % He & Helium\\
    % C & Carbon\\
    % O & Oxygen\\
    % Ne & Neon\\
    % Mg & Magnesium\\
    % Si & Silicon\\
    % Fe & Iron\\
    % Ni & Nickel\\
    IMF & Initial mass function, that is distribution of initial
          masses of stars\\
    CSM & Circumstellar material
\end{tabular}
\end{glossary}

\begin{abstract}[Abstract]
  Very massive stars are radiation pressure dominated. Before running
  out of viable nuclear fuel, they can reach a thermodynamic state
  where electron-positron pair-production robs them of radiation
  support, triggering their collapse. Thermonuclear explosion(s) in
  the core ensue. These have long been predicted to result in either
  repeated episodic mass loss (pulsational pair instability), which
  reduces the mass available to eventually form a black hole, or, if
  sufficient energy is generated, the complete unbinding of all
  stellar material in one single explosive episode (pair instability
  supernova), which leaves behind no black hole. Despite theoretical
  agreement among modelers, the wide variety of predicted signatures
  and the rarity of very high-mass stellar progenitors have so far
  resulted in a lack of observational confirmation. Nevertheless,
  because of the impact of pair instability evolution on black hole
  masses relevant to gravitational-wave astronomy, as well as the
  present and upcoming expanded capabilities of time-domain astronomy
  and high redshift spectroscopy, interest in these explosion remains
  high. We review the current understanding of pair instability
  evolution, with particular emphasis on the known uncertainties. We
  also summarize the existing claimed electromagnetic counterparts and
  discuss prospects for future direct and indirect searches.
\end{abstract}

\begin{BoxTypeA}[]{Key points}
\begin{itemize}
\item Very massive, radiation pressure dominated stars encounter the
  pair-instability near the ends of their lives, but before they run out of nuclear fuel.
\item The pair instability converts radiation energy density into
  rest-mass of the $e^\pm$ pair, softening the radiation-pressure
  dominated equation of state and causing a collapse. Heating due to gravitational
  collapse triggers a thermonuclear explosion.
\item In pulsational pair instability supernovae (P-PISN), the
  explosion causes the (possibly repeated) ejection of layers of mass, while
  in pair instability supernovae (PISN) it completely destroys the
  star, leaving no compact remnant.
\item The observed light from P-PISNe is powered by the interaction
  between ejected shells, while the luminosity of PISNe is
    powered by the radioactive decay of large amount of
  radioactive nickel ($^{56}$Ni).
\item Despite theoretical agreement among modelers, GW detection of
  BHs ``in the PISN mass gap'' have occured, while the bright SNe associated with
  pair-instability have so far eluded detection.
\item Theoretical studies continue to explore uncertainties in the
  input physics, stellar modeling, and interpretation of GW and
  electromagnetic transients. In the near future, multiple
  observations may help solve the apparent discrepancy between
  theoretical predictions and observations.
\end{itemize}
\end{BoxTypeA}

\section{Introduction}

The final fate of massive stars (initially $\gtrsim 7.5\,M_\odot$,
e.g.,~\citealt{doherty:15, poelarends:17}) is a long-standing
problem in astrophysics. For millennia \citep[e.g.,][]{xi:55,
  clark:77, stephenson:02, zhou:18}, we have been able to observe
spectacular transients, called supernovae (SNe), which occur when a
massive star's core exhausts viable nuclear fuel. This leads to the collapse
of the iron (Fe) core (or, at the lower mass end, the oxygen-neon core) under its own
gravity until it ``bounces'' at super-nuclear densities, triggering a
shock wave that \emph{possibly} blows up the star\footnote{These explosions are distinct from thermonuclear SNe, typically
observed as Type Ia \citep[e.g.,][for a recent review]{liu:23}, which
correspond to the thermonuclear explosion of a white dwarf in a
low-mass binary system. % that was
% the evolutionary endpoint of a low-mass binary system, leaving no
% compact remnant. \mr{there may be a remnant in SN .Ia}
}. These events mark
the birth of compact objects (neutron stars, NS, or black holes, BHs).
Despite being routinely observed, they still pose theoretical
challenges (see \citealt{janka:12, burrows:21, soker:24} for reviews).

The situation is reversed for even more massive stars, those
\emph{ending} their lives with helium (He) core masses
$M_\mathrm{He}\gtrsim 50\,M_\odot$, roughly corresponding to initial
masses $\gtrsim 70-100\,M_\odot$ \citep[e.g.,][]{barkat:67, rakavy:67,
  fraley:68, bond:84, woosley:17}. Before running out of viable
nuclear fuel, these very massive and radiation-pressure dominated
stars have cores that are hot enough and have low-enough density to
trigger an instability that follows from converting radiation energy
density into rest-mass of electron-position pairs ($e^\pm$), robbing
the core of some of its pressure support. This is the so called ``pair
instability'', which causes collapse and a thermonuclear explosion in
very massive stars. %  \citep[e.g.,][]{fowler:64, rakavy:67, barkat:67,
  % woosley:17}.
There is theoretical consensus on how these stars end their
lives, but unambiguous observational evidence is still lacking. The likely
reasons are the intrinsic rarity of their progenitors -- heavily
disfavored by the stellar initial mass function
\citep[e.g.,][]{salpeter:55, kroupa:01, schneider:18} and the large
range of predicted observational signatures overlapping with other
transients \citep[e.g.,][]{woosley:07, woosley:17, smith:17,
  renzo:20:ppi_csm, woosley:21}. The end of the evolution of these
stars % and their final fate
is the topic of this chapter.

Interest in the final fate of these extremely massive stars is ever
growing. They are thought to be the progenitors of the most massive
stellar-mass BHs, bridging the gap between the stellar-mass BHs now
routinely detected through gravitational waves \citep[GW,
e.g.,][]{first_detection, GWTC1, GWTC2, GWTC3}, and intermediate-mass
BHs ($\gtrsim 100M_{\odot}$, \citealt{mehta:22}). Improvements in
time-domain capabilities (sky coverage, cadence, depth) are revealing
rare transients that should include the death events of extremely rare
very massive stars, and combined astrometric, photometric, and most
importantly spectroscopic observations enable the search for
nucleosynthetic signatures of the death of these stars
\citep[e.g.,][]{aoki:14, xing:23}. Finally, numerical development of
stellar evolution codes has recently allowed the self-consistent
exploration of the entire mass range \citep[e.g.,][]{woosley:07,
  woosley:17, woosley:19, marchant:19, farmer:19, leung:19,
  renzo:20:ppi_csm, farag:22}. This has enabled new theoretical
insight on the evolution of very massive stars, and
has raised new questions in stellar, nuclear, and particle physics
(including beyond-standard model physics).

We first review the theoretical understanding of the death of very
massive stars (\Secref{sec:intro_theory}). We discuss the expected
observational signatures (\Secref{sec:obs_predictions}), and we
summarize claims in the literature of candidates for direct detections
of optical transients caused by the pair instability
(\Secref{sec:obs_EM}). We discuss indirect evidence
(\Secref{sec:indirect}), including GWs (\Secref{sec:GW_obs}), before
highlighting the known open questions (\Secref{sec:known_unknowns}).
We conclude suggesting future prospects in \Secref{sec:conclusions}.

\section{Theoretical overview of pair instability}\label{sec:intro_theory}

\subsection{Microphysics}\label{sec:microphysics_theory}

Here, we discuss the microphysics that determines the death of stars
susceptible to the pair instability. We defer a quantitative
discussion of the minimum mass needed for this process to
\Secref{sec:BH_mass_gap}. The cores of these stars experience a
(potentially destructive) instability before running out of viable
nuclear fuel \citep{barkat:67, rakavy:67}, usually when the core
composition is oxygen-dominated. Because of their large mass, these
stars are radiation pressure dominated,
\begin{equation}
  \label{eq:pressure}
  P_\mathrm{tot} = P_\mathrm{gas}+P_\mathrm{rad} \simeq P_\mathrm{rad}
  % = \frac{1}{3} u_\mathrm{rad} \equiv
  =  \frac{1}{3}aT^4 \ \ ,
\end{equation}
where % $u_\mathrm{rad}$ is the radiation energy density,
$T$ is the local temperature, $a=4\sigma/c$ is the radiation constant
($\sigma$ is the Stefan-Boltzmann constant and $c$ the speed of
light). This is because
virial % comes from vir, latin for "force", no capitalization!
equilibrium imposes $T\propto M$, with $M$ total mass of the star and
$T$ average temperature, and the radiation pressure
$P_\mathrm{rad}\propto T^4$ grows with temperature much faster than
the thermal gas pressure $P_\mathrm{gas}\propto T$.

The hydrostatic evolution of these stars brings them into a
thermodynamic regime where the production of electron-position pairs
($e^\pm$) occurs \citep{breit:34, fowler:64},
\begin{equation}
  \label{eq:pair_production}
  \gamma\,\gamma \rightarrow e^+\,e^- \ \ .
\end{equation}
Each $e^\pm$ pair produced may then annihilate producing either a  neutrino-antineutrino pair, which stream out of the star carrying away
their energy, or into a pair of photons, which on average have lower
energy than the original photons\footnote{The electron or positron for
  the annihilation is usually not the same as the one produced in the original
  pair.}. This process can in principle happen in any star, since it
only requires the interaction of photons with total energy in excess
of the rest energy of the $e^\pm$ pair
($E_\gamma\gtrsim 2 m_{e}c^{2}$, where $m_e$ is the electron mass).
Normally, stellar interiors are in thermal equilibrium, meaning the
photons follow a black-body distribution with an exponentially
decaying tail at high-energy that yields a small, but non-zero, probability
of having such high energy photons: pair-production alone does not
necessarily result in an instability. However, when this occurs in a
star whose hydrostatic equilibrium depends on radiation pressure, it
can lead to a catastrophic instability: the production of $e^\pm$
pairs effectively consumes radiation energy density. %$u_\mathrm{rad}$
% that goes into the rest-energy of the $e^\pm$ pair.
This %effectively
softens the equation of state, causing a \emph{local} thermal
instability in the star. We discuss the \emph{global} response of the
star (if any) in \Secref{sec:evol_theory}.

\begin{figure}[htbp]
  \centering
  \includegraphics[width=0.5\textwidth]{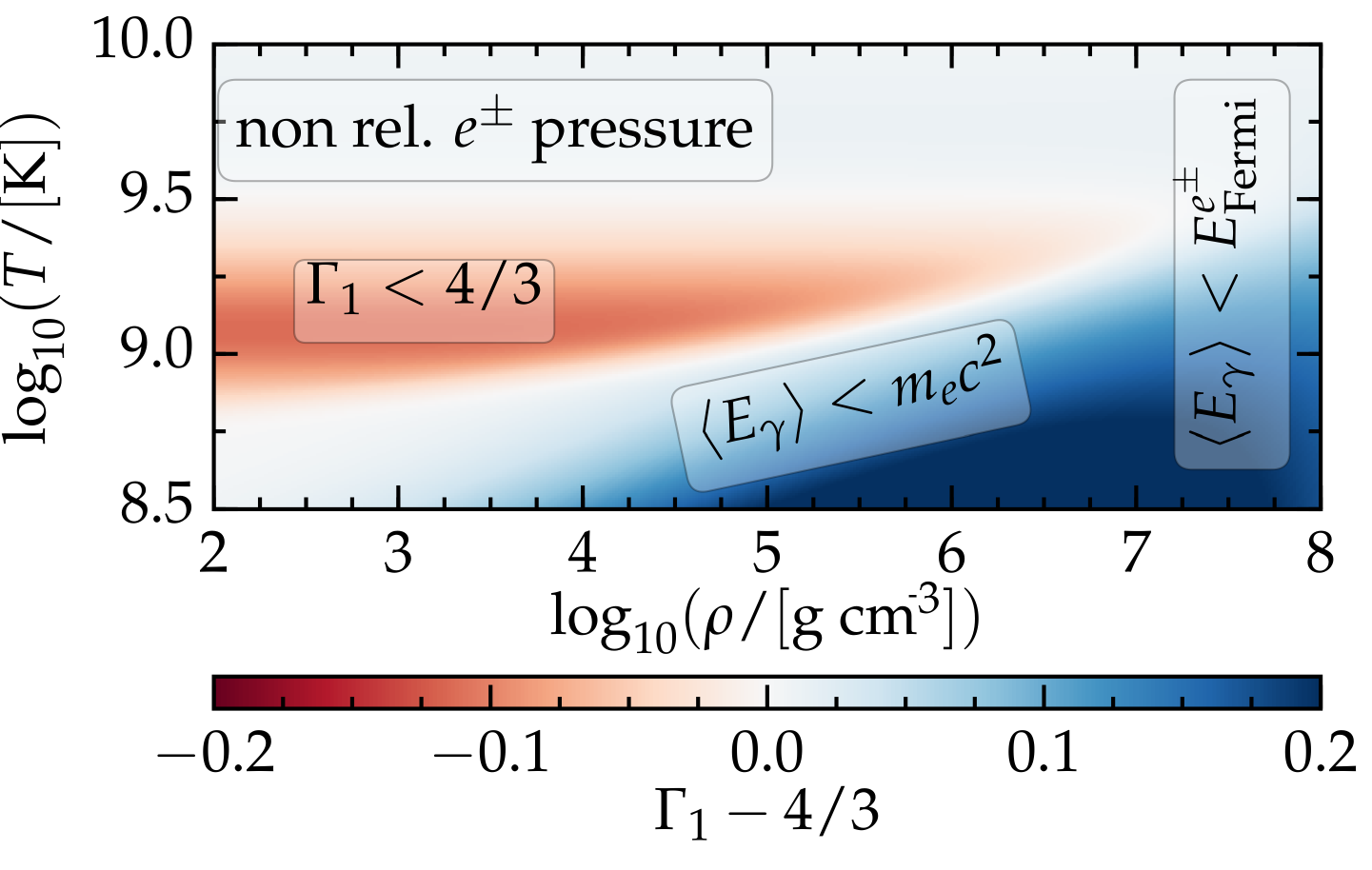}
  \caption{First adiabatic index as a function of temperature and
    density for a composition typical of a PISN progenitor (mixture
    dominated by carbon and oxygen). In the red
    region $\Gamma_1<4/3$ resulting in a thermal instability with
    runaway production of $e^\pm$. From
    \cite{renzo:19:phd}.}
  \label{fig:instability}
\end{figure}

\Figref{fig:instability} shows first adiabatic index
$\Gamma_1 = \frac{\partial \ln T}{\partial \ln \rho}\lvert_{s}$ (where
$s$ is the specific entropy) on the temperature-density ($T-\rho$)
plane the for a composition representative of an evolved very massive
star, that is mixture of carbon (C) and oxygen (O) with $\lesssim0.1\%$ of
other primordial metals. Classically, the adiabatic index $\Gamma_1$ is
used to discuss the stability of a stellar structure \citep[requiring
$\Gamma_1>4/3$, e.g.,][]{kippenhahn:13}, and \Figref{fig:instability}
shows the local value of $\Gamma_1$. Stellar layers entering the red
regions, where $\Gamma_1<4/3$, are destabilized by the production of
$e^\pm$ pairs, which occurs around $T\gtrsim10^9\,\mathrm{K}$
corresponding to photon energies of the order of
$m_ec^2\simeq0.511\,\mathrm{MeV}$ (the pairs are produced by photons
in the tail of the black body distribution). \Figref{fig:instability}
also annotates the microphysical mechanisms that delimit the region of
local instability (see also \Figref{fig:gamma_tracks}). At
sufficiently high temperatures ($T\gtrsim 10^{9.5}\,\mathrm{K}$), the
instability ``chokes'' itself: $e^\pm$ pairs are produced so fast that
their Fermi levels are filled, and the pairs themselves start
supporting the star with their non-relativistic degeneracy pressure,
bringing back $\Gamma_1\gtrsim 4/3$ \citep[e.g.,][]{kippenhahn:13}.
Because of the energy requirement $E_\gamma\gtrsim 2m_ec^2$,
pair-production is a threshold process and most $e^\pm$ pairs are
produced with negligible kinetic energy making them non-relativistic.

At sufficiently high densities
$\rho\gtrsim 10^7\,\mathrm{g\ cm^{-3}}$, the $e^\pm$ are also
degenerate \citep[e.g.,][]{fraley:68}, and for pair-production to
occur, the photon energy must exceed not only the rest-energy of the
pair, but must also be sufficient to provide kinetic energy in excess
of their Fermi energy \citep{zeldovich:99}. This extra energy
requirement decreases the rate of pair production and prevents it from
triggering an instability. Moreover --- and possibly more importantly
--- at higher densities, the contribution of the gas pressure becomes
increasingly important, protecting the star from catastrophic outcomes
that would otherwise occur due to the drop in radiation pressure
caused by pair-production. If and when lower mass stars reach inner
temperatures sufficient for pair-production, the combination of these
two effects prevents pair-production from triggering global
instabilities.

At low temperatures, the mean photon energy is insufficient to produce
pairs at rest, the precise shape of the boundary is determined by the
requirement that the mean-free-path for photon-photon
interactions % $\lambda_{\gamma\gamma}$
has to be smaller that the mean-free-path % $\lambda_{\gamma e}$
for photon interactions with an electron (which are the dominant
source of opacity in this regime).% : only if
% $\lambda_{\gamma\gamma} < \lambda_{\gamma e}$ pair-production can
% become an instability.

\subsection{Stellar evolution through pair instability}\label{sec:evol_theory}

\begin{figure}[htbp]
  \centering
  \includegraphics[width=0.75\textwidth]{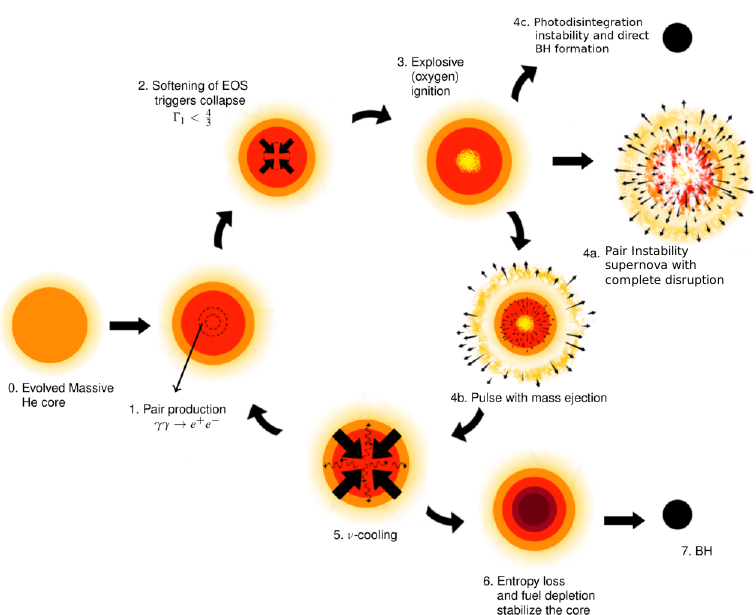}
  \caption{Cartoon representing the core evolution through (P)PISN,
    from \cite{renzo:20:ppi_csm}. Step 4 represents the trifurcation
    point depending on the mass of the core: 4a~represents the full
    disruption in a PISN occurring when the energy release by the
    thermonuclear explosion exceeds the binding energy of the star.
    For lower masses, the energy release can only unbind the outer
    layers, resulting in a pulse, and the cycle can repeat, producing
    a P-PISN (4b). For higher masses, the energy released is used to
    photo-disintegrate nuclei instead of accelerating gas, resulting in
    collapse to a BH (4c).}
  \label{fig:cartoon}
\end{figure}

\Figref{fig:cartoon} illustrates the phases of the stellar response to
the instability described in \Secref{sec:microphysics_theory} caused
by the production of $e^\pm$ pairs. Step 0 starts with a core
sufficiently massive to be radiation pressure-dominated and encounter
the instability, typically when its core composition is still
dominated by viable nuclear fuel -- C, O, possibly Silicon (Si) group
elements \citep{marchant:19}.

The instability starts at Step 1, where pair-production removes
photons that were providing pressure support to the star. Because of
the drop in pressure support, the star will contract. This contraction
will increase its temperature, and thus also the average energy of
photons in its interior $\langle E_\gamma\rangle$. Consequently the
amount of photons that can do pair-production also increases. This
causes a runaway production of pairs that hastens the collapse.

As the star collapses and increases its temperature, the nuclear fuel
will ignite (step 3). This ignition is necessarily explosive, since
the star is out of thermal equilibrium and the nuclear reaction rates
are strongly temperature dependent: thus pair-instability leads to
thermonuclear explosions in massive stars, and can result in three
different outcomes depending on the mass of the star, which determines
the ratio between the energy released by the explosive burning and the
binding energy of the stellar gas.

The classical pair instability SNe (PISN, step 4a), theorized for the
first time in the 1960s \citep[e.g.,][]{barkat:67,
  rakavy:67,fraley:68, glatzel:85, fryer:01, woosley:07,
  chatzopoulos:12, chatzopoulos:13, woosley:17, marchant:19, leung:19,
  farmer:19, renzo:20:conv_ppi, renzo:20:ppi_csm, farag:22, umeda:24}
occurs when the nuclear burning releases more energy than the entire
binding energy of the star. In these cases, the explosions leaves no
compact remnant and all the stellar material is returned to the host
galaxy: PISN lead to full disruption of the star and no BH formation.

For slightly lower masses, the central temperature reached (and
consequently, the nuclear burning rate) are lower: this results in
less energetic explosions that fail to completely unbind the star. In
these cases, the energy deposition from explosive burning triggers a
pulse that propagates through the star, steepens into a shock wave as
it propagates through the lower-density outer layers. This pulse may
unbind some material, in a so-called pulsational pair-instability SN
(P-PISN, \citealt{barkat:67, woosley:07, chatzopoulos:12,
  chatzopoulos:13, yoshida:16, woosley:17, leung:19, marchant:19,
  farmer:19, farmer:20, renzo:20:ppi_csm, renzo:20:conv_ppi,
  farag:22}), at step 4b. These are not terminal events in the life of
a star, but can appear as bright transients \citep[e.g.,][see also
\Secref{sec:EM_obs}]{woosley:07, woosley:17,lunnan:18, gomez:19,
  woosley:22}. Even if not unbinding the entire star, the explosion
causes an expansion of the core, lowering its temperature and density
and pulling it out of the instability region
(cf.~\Figref{fig:instability}). The star then relaxes on a thermal
timescale, which in this regime is typically determined by neutrino
cooling \citep[e.g.,][]{barkat:67, fraley:68}, although the most
energetic pulses may drive the core to sufficiently low densities such
that neutrino cooling is shut off, and photons (and convection) return
to be the main energy carriers \citep[e.g.,][]{woosley:17,
  marchant:19, renzo:20:ppi_csm}. This relaxation timescale can vary
widely from days to $10^5\,\mathrm{yrs}$ \citep[e.g.,][]{woosley:17,
renzo:20:ppi_csm}. After this relaxation phase, the star may encounter the instability
again, looping through the cycle, until (\emph{i}) the consumption of
viable nuclear fuel at step 3, (\emph{ii}) the loss of mass at step
4b, and (\emph{iii}) the loss of entropy during the cooling phase at
step 5 stabilize the core, making it avoid the pair-production
instability. At this point, normal stellar evolution resumes,
finishing to burn in hydrostatic equilibrium what nuclear fuel is left
in the core, ultimately producing an Fe core-collapse
event (which may or may not be associated to an electromagnetic
transient). In the mass regime of interest here, it is expected that
such core-collapse would result in BH formation -- but its final mass
is limited by the amount of mass previously lost during pulses
\citep[e.g.,][]{woosley:02, woosley:17, marchant:19, farmer:19,
  farmer:20, renzo:20:ppi_csm, woosley:21, farag:22} and possibly in
an explosion associated to the final core-collapse (though the ejecta
are expected to be $\lesssim3.5\,M_\odot$, \citealt{powell:21,
  rahman:22} but is, however, sensitive to uncertainties in
core-collapse physics, see also \citealt{hendriks:23}).

For stars even more massive than PISN progenitors, a second
instability occurs, reffered to as the photodisintegration instability
\citep[e.g.,][]{bond:84, fryer:01, heger:03}. In fact, the rate of
photodisintegration, when sufficiently energetic photons are
available, scales as $\propto \rho$, while the rate of nuclear
reactions dominating the energy release (initially
$^{16}\mathrm{O}+^{16}\mathrm{O}$, \citealt{dessart:13, farmer:19})
scale like $\propto \rho^2$, and the latter can be lower than the
former for sufficiently low densities. Thus, in extremely massive and
thus low density stars, the energy released by the thermonuclear
explosion is used to photo-disintegrate nuclei (unburned C, O, Si, and
the products of the thermonuclear explosions itself). This means that
all the energy is consumed to shuffle nucleons in and
out of nuclei, instead of becoming bulk kinetic energy of the stellar
gas in an explosion. The photodisintegration instability therefore
cancels out the dynamical effect of the pair-instability, and results
in the collapse of the star into a massive BH.

\begin{figure}[htbp]
  \centering
  \includegraphics[width=0.5\textwidth]{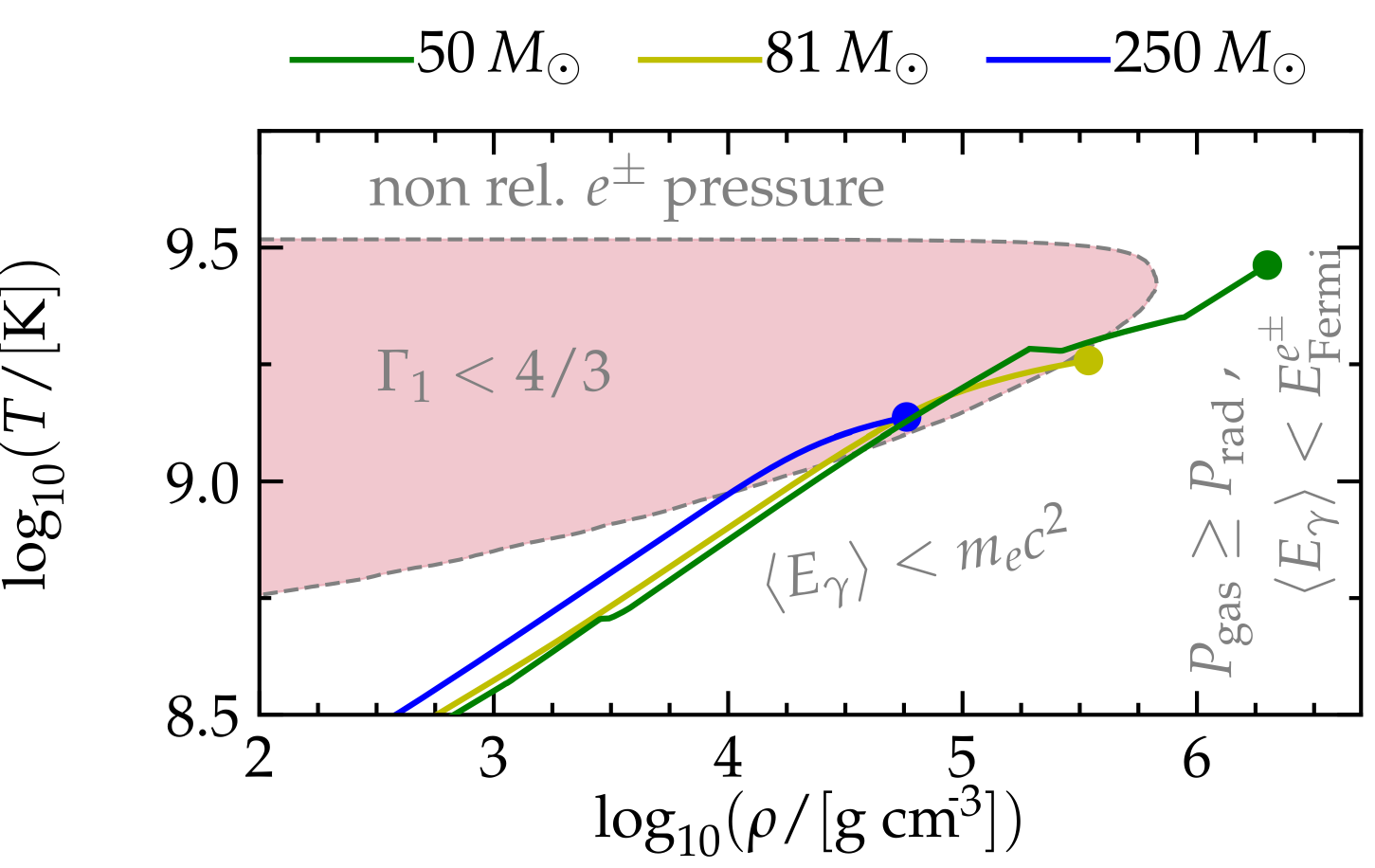}
  \caption{$T-\rho$ profile for He cores modified from \cite{renzo:20:ppi_csm}
    at the moment they approach the instability, that is when
    $\langle \Gamma_1 \rangle \simeq 4/3$. Models where the center is
    unstable ($\Gamma_{1,c} <4/3$), experience a PISN with full
    disruption (yellow profile, cf.~step 4a in \Figref{fig:cartoon}).
    Models where the center is stable ($\Gamma_{1,c} >4/3$),
    experience a P-PISN (green profile, cf.~step 4b in
    \Figref{fig:cartoon}). Extremely massive models with very low
    central densities ($\rho\lesssim 10^5\,\mathrm{g \ cm^{-3}}$)
    experience the photodisintegration instability and collapse to BHs
    (blue profile, cf.~step 4c in \Figref{fig:cartoon}). The colors
    label the total initial He core mass, no H envelope was included
    in the models of \cite{renzo:20:ppi_csm}.}
  \label{fig:gamma_tracks}
\end{figure}

Which of the three branches at step 4. a star takes is primarily
determined by how much mass has low enough entropy and high enough
temperature to be involved in the thermonuclear explosion, which is
well approximated by the CO core mass \citep{farmer:19, farmer:20}.
This, in turn, is determined by the evolutionary history of the He
core \citep[e.g.,][]{woosley:19, farmer:19, farmer:20,
  renzo:20:conv_ppi, renzo:20:ppi_csm, woosley:21} and thus the
initial mass of the star and its evolution, including mass loss or
interaction in binaries \citep[e.g.,][]{marchant:19, vanson:20,
  laplace:21, renzo:23, hendriks:23, schneider:24} and clusters
\citep[e.g.,][]{spera:19, dicarlo:19, dicarlo:20a, dicarlo:20b,
  kremer:20, mapelli:20, costa:22, ballone:23}.

\cite{renzo:20:ppi_csm} proposed a
criterion to discriminate between steps 4a and 4b illustrated in
\Figref{fig:gamma_tracks} and based on the pressure-weighted average
$\Gamma_1$,
\begin{equation}
  \label{eq:P_weighted_gamma1}
  \langle\Gamma_1\rangle = \frac{\int \Gamma_1P\,d^3r}{\int P\,d^3r} \equiv \frac{\int \Gamma_1 \frac{P}{\rho}\,dm}{\int
    \frac{P}{\rho}\,dm} \ \ ,
\end{equation}
where $r$ is the radial coordinate and $m$ is the mass coordinate
throughout the star, and the value of $\Gamma_{1,c} = \Gamma_1(r=0)$
in the center of the star. The pressure-weighting in
\Eqref{eq:P_weighted_gamma1} is motivated by the fact that we want to
predict the dynamical response of the star \citep{stothers:99}, and it
also naturally up-weights the high-pressure core region where the
instability occurs. Stars approaching the instability
($\langle \Gamma_1\rangle \simeq 4/3$) with a stable center
($\Gamma_{1,c}>4/3$) experience P-PISN but finally form a BH, while
stars unstable also in their center ($\Gamma_{1,c}<4/3$) experience full disruption in a
PISN. Provided a stellar profile at the onset of the instability, one
can use this criterion to decide its fate without computing the
dynamical phase of evolution. Models experiencing the
photodisintegration instability can be identified because of their
extremely low central density, needed for photodisintegrations.
Nevertheless, computing the evolution of stars through P-PISN pulses
(and their final core-collapse, \citealt{powell:21, rahman:22})
remains necessary to determine the final BH masses.

\subsection{Which stars undergo pair instability?}\label{sec:mass_ranges}

The processes triggered by pair-instability are all happening deep in
the CO core. The mass of this CO core, as opposed to the star's
initial mass, is the better metric to decide whether a star goes
(P)PISN or not \citep[e.g.,][]{farmer:19}. However, many uncertain
physical processes in the evolution of massive and very massive stars
(see \Secref{sec:known_unknowns}), together with the likely occurrence
of binary interactions and mergers \citep[e.g.,][]{vignagomez:19,
  dicarlo:19, dicarlo:20a, dicarlo:20b, renzo:20:merger, costa:22,
  ballone:23} make it impossible to have a robust and
one-to-one mapping of initial masses with CO core masses
\citep[e.g.,][]{zapartas:21}. Consequently, any number quoted in this
section is affected by large uncertainties and subject to changes as
the input (e.g., nuclear reaction rates) and stellar (e.g., wind mass
loss rates, convective boundary mixing) physics are updated (see
\Secref{sec:known_unknowns}).

Roughly speaking, for (P)PISN to remove significant amount of mass
requires CO core masses of $M_{\rm CO}\gtrsim 45\,M_\odot$, roughly
corresponding to $M_\mathrm{He}\gtrsim 50\,M_\odot$. These values are
not the initial values, but rather the values at the time of the onset
of the instability (when $\langle\Gamma_1\rangle \lesssim 4/3$). Very
roughly, stars with initial total (H-rich) masses of
$\sim{}70-100\,M_\odot$ at sufficiently low metallicity may develop
and retain sufficiently massive He and CO cores to trigger (P)PISN
\citep[e.g.,][]{woosley:02, woosley:07, woosley:19, yusof:13,
  umeda:20, yusof:22}. Stars with lower CO core masses
($M_\mathrm{CO}\gtrsim30\,M_\odot$) can also experience ``mild''
P-PISN which does not result in large mass ejections
\citep[e.g.,][]{barkat:67, woosley:17, woosley:19, renzo:20:ppi_csm}
but may still be detectable (see \Secref{sec:pulse_types}).

In terms of metallicity, because of the requirement of retaining a
large He and CO core until the end of the evolution, (P)PISN is often
discussed in the context of metal-free population III stars
\citep[e.g.,][]{heger:03, whalen:13, umeda:20, nagele:22}. Based on
hydrostatic models, \cite{langer:07} inferred an upper-limit of
$Z\lesssim Z_\odot/3$ for the maximum metallicity allowing for
(P)PISN. This may need to be adjusted to somewhat higher metallicity,
however, due to modern reductions in mass-loss rate prescriptions
\citep[e.g.,][]{smith:14}. This upper-limit may also be exceeded by stellar
mergers producing large CO cores late in the evolution, leaving no
time for the star to lose mass enough mass post-merger
\citep[e.g.,][]{vignagomez:19, renzo:20:merger, costa:22}, or because
of magnetic fields funneling back the stellar wind onto the star,
preventing efficient wind mass loss \citep[e.g.,][]{georgy:17,
  keszthelyi:19}.

\subsection{The theorized PISN BH mass gap}\label{sec:BH_mass_gap}

\begin{figure}[bp]
  \centering
  \includegraphics[width=0.75\textwidth]{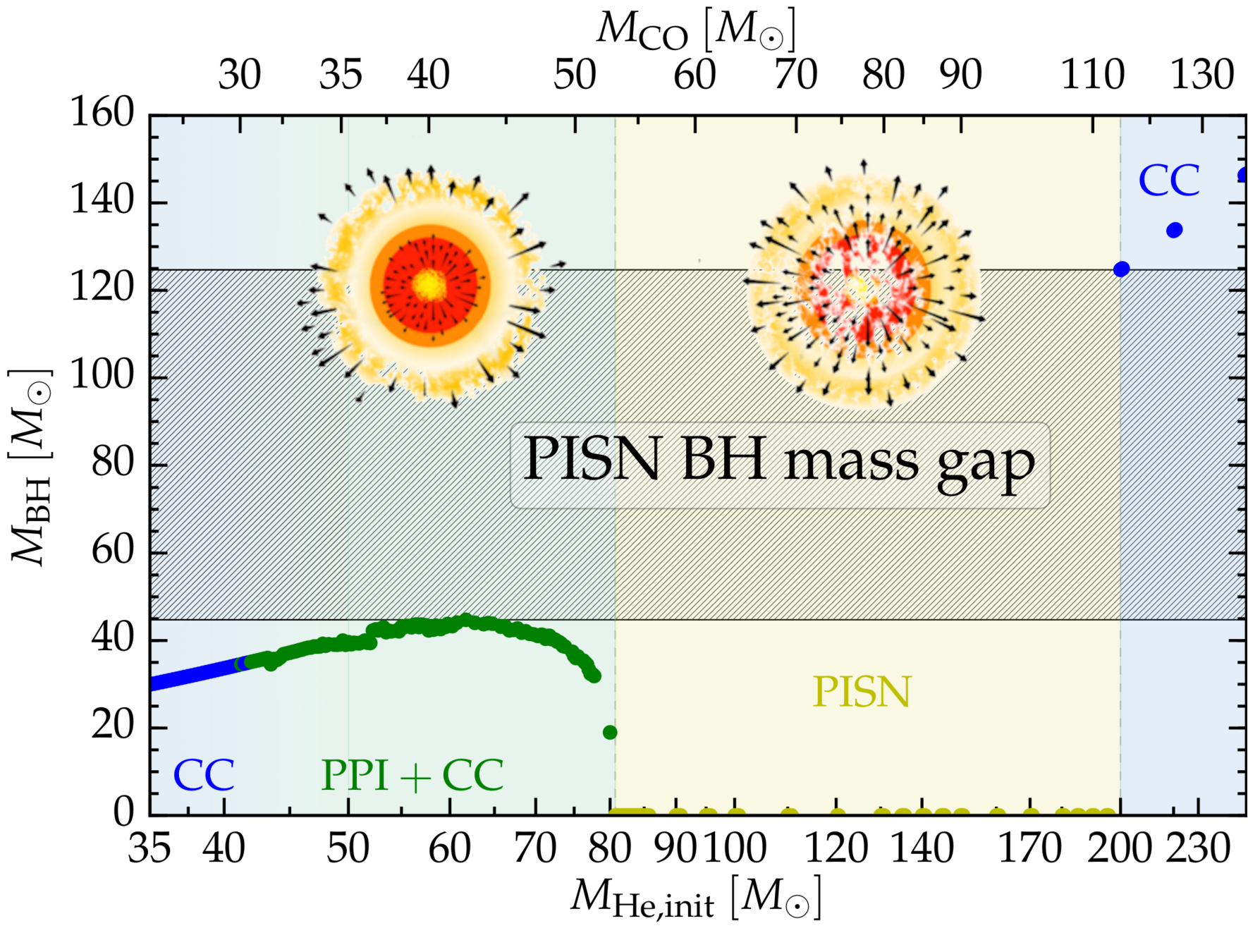}
  \caption{Schematic illustration of the BH mass gap modified from
    \cite{renzo:20:ppi_csm}. The BH mass gap is on the y-axis (black
    hatch), subject to large uncertainties (see
    \Secref{sec:known_unknowns}). The most up-to-date values based on
    the evolution of He cores are from \cite{farag:22} and
    \cite{mehta:22}, with the lower edge at masses higher than
    $\sim45\,M_\odot$, which is an old value that should be disfavored
    based on more recent predictions -- see also
    \Figref{fig:mehta_gap} and text. The background colors indicate
    the predicted fate of stars: blue for core-collapse (CC), green
    for P-PISN followed by core-collapse (PPI+CC), yellow for PISN.
    The bottom x-axis shows the initial He core mass for the stellar
    models, and the top x-axis shows the maximum CO core mass
    throughout the evolution, which is a more direct parameter to
    determine the fate of a star. Models from \cite{renzo:20:ppi_csm}
    only include the He core.}
  \label{fig:PISN_BH_mass_gap}
\end{figure}

While the idea that no BHs of a certain mass would form from the
collapse of stars because of PISN was already proposed in the 1960s,
until the 21$^\mathrm{st}$ century, stellar evolution calculations of
the entire mass-range of interest remained challenging. This is
because during P-PISN the stars experience dynamical phases of
evolution (step 4b in \Figref{fig:cartoon}) with hydrostatic phases in
between (steps 5-1-2 post-pulse) -- two regimes that require different
approximations and numerical treatments, therefore posing numerical
challenges. With the exception of the closed-source code KEPLER
(\citealt{weaver:78} with the relevant updates described in
\citealt{woosley:02, woosley:07, woosley:17, woosley:19, woosley:21}),
% early \N{(hmmm... would you really call MESA "early approaches" -
%   maybe say the first modern approaches or something? early sounds
%   like the 1960s. you are younger than me, i guess...)}
the typical approach was to combine stellar evolution calculations
(e.g., with codes such as MESA, \citealt{paxton:11, paxton:13,
  paxton:15, paxton:18, paxton:19, jermyn:23} -- or HOSHI,
\citealt{yamada:97}) with analytic considerations or hydrodynamical
calculations with a different code (e.g., FLASH, \citealt{fryxell:00}
-- or \texttt{v1D} \citealt{livne:93}) for the dynamical phase of the
pulses \citep[e.g.,][]{chatzopoulos:12, chatzopoulos:13, dessart:13,
  takahashi:16, kozyreva:17, takahashi:18, umeda:20, nagele:22,
  umeda:24}. While this approach allows for the study of the
hydrodynamics of one pulse, and is thus suitable for studying PISN, it
makes it hard to follow the evolution through multiple pulses.
Nevertheless, following a model through multiple pulses is needed to
estimate the final, post-P-PISN, BH masses. Other stellar evolution
codes (like PARSEC, see e.g.,~\citealt{costa:21, costa:22, costa:23},
and GENEC, see e.g.,~\citealt{eggenberger:08, yusof:13, kozyreva:17,
  yusof:22}) only evolve stellar models until core C depletion (before
the pair-instability may occur) and assess the presumed fate based
only on the CO core mass, predicting upper-limits for the lower-edge
of the theoretical PISN BH mass gap that neglect the mass lost during
P-PISN pulses.

The detection of GWs revived the interest in the BH mass gap, leading
to the development of open-source and community-driven codes capable
of alternating between dynamical pulses and hydrostatic phases of
relaxation \citep[e.g.,][]{paxton:18, leung:19, marchant:19,
  farmer:19, farmer:20, renzo:20:ppi_csm, renzo:20:conv_ppi,
  farag:22}. This enabled the exploration of theoretical uncertainties
on the location of the gap \citep{farmer:19,
  farmer:20,renzo:20:conv_ppi, mehta:22, farag:22}, and even the
inclusion of beyond-standard-model physics (see
\Secref{sec:beyond_SM}). The lower edge of the gap has received far
more attention, partly because some estimates of its location overlap
with BH masses detected through GWs, and partly because no star
simultaneously sufficiently massive and metal poor to end its life
above the gap is known (see also, e.g., \citealt{dekoter:97,
  crowther:10, crowther:16, woosley:17, renzo:19vfts682}). Predictions
for GW detections further depend on the dynamical
\citep[e.g.,][]{dicarlo:19, dicarlo:20a, dicarlo:20b,renzo:20:merger,
  kremer:20, mapelli:20} or binary interactions
\citep[e.g.,][]{spera:19, vignagomez:19, marchant:19, vanson:20,
  hendriks:23} between the stellar progenitors of the merging BHs,
which introduce further uncertainties and are under very active
investigation.

Since they can cause full disruption of massive stars leaving no
compact remnant, PISNe are predicted to carve a ``gap'' in the BH mass
distribution \citep[e.g.,][]{rakavy:67, fraley:68, woosley:02,
  chatzopoulos:12, woosley:17, takahashi:18, marchant:19, farmer:19,
  farmer:20, renzo:20:ppi_csm, woosley:21, farag:22}, illustrated on
the y-axis of \Figref{fig:PISN_BH_mass_gap}. The lower edge of this
gap is determined by total the amount of mass lost during P-PISN
(which are expected to form a BH after the pulses,
\citealt{woosley:17, renzo:20:ppi_csm, powell:21, rahman:22}). Then,
for the entire mass range where PISNe occur, no BHs form (yellow
region in \Figref{fig:PISN_BH_mass_gap}). The upper edge of the gap is
determined by the onset of the photodisintegration instability
\citep[e.g.,][see also \Secref{sec:evol_theory}]{bond:84}, which
produced massive BHs. This is sometimes referred to as the ``second''
mass gap, to distinguish it from the (highly debated) gap between the
most massive NS and the least massive BHs (e.g., \citealt{ozel:10,
  farr:11}, but see also e.g., \citealt{wyrzykowski:20, GW190814}).

% In \Figref{fig:PISN_BH_mass_gap}, the bottom x-axis shows the initial
% He core mass simulated -- \citealt{renzo:20:ppi_csm} did not include
% H-rich envelopes, assumed to be lost to either winds or binary
% interactions long before the (P)PISN -- while the top x-axis shows the CO core
% mass (maximum pre-pulse value), which is a better proxy for the amount
% of mass involved in the instability \cite[e.g.,][]{farmer:19}.
The values on the axes of \Figref{fig:PISN_BH_mass_gap} are subject to
many uncertainties \citep[e.g.,][see also
\Secref{sec:known_unknowns}]{farmer:19, farmer:20, renzo:20:conv_ppi,
  woosley:21, mehta:22, farag:22}. %, which also impact the resulting BH masses on
% the y-axis.
The BH masses are estimated assuming full collapse of the matter
remaining post-pulse (see also \Secref{sec:fate_envelope}). The onset
of P-PISN is marked by a smooth transition between the blue on the
leftmost part of \Figref{fig:PISN_BH_mass_gap} and the adjacent green
region: this is because what exactly is identified as a P-PISN pulse
depends on the target observable \citep{renzo:20:ppi_csm}. These
results are also affected by several known uncertainties in the input
physics, and algorithmic representation of relevant physical processes
in spherical symmetry (see
\Secref{sec:known_unknowns}). % Roughly speaking,
% significant mass loss during P-PISN requires CO core masses
% $M_\mathrm{CO}\gtrsim 45\,M_\odot$ corresponding to
% $M_\mathrm{He}\gtrsim 50\,M_\odot$ at the onset of the pulses, meaning
% after most of the stellar life has elapsed and wind mass loss has
% decreased the total and core masses \citep[e.g.,][]{vink:15,
%   sabhahit:23}.

% The predicted location of the theorized PISN BH mass gap (on the y
% axis in \Figref{fig:PISN_BH_mass_gap}) is sensitive to both input
% physics and numerical approximations (see
% \Secref{sec:known_unknowns}), including especially the fate of the
% hydrogen-rich envelope if present at the onset of the pulses (see
% \Secref{sec:fate_envelope}).

The most updated values for the theorized location of the PISN BH mass
gap (based on He core models) are from \cite{farag:22}, who found for
the lower edge of the gap
$\max(M_\mathrm{BH}) =69^{+34}_{-18}\,M_\odot$, and \cite{mehta:22}
who found for the upper edge of the gap
$\min(M_\mathrm{BH})=139^{+30}_{-14}\,M_\odot$ (see also
\Figref{fig:mehta_gap}). The ranges quoted here bracket uncertainties
in certain nuclear reaction rates (see also \Secref{sec:unknown_nuc}).
These numbers are in broad agreement with the values reported by many
other authors \citep[e.g.,][]{woosley:02, woosley:17, woosley:19,
  marchant:19, farmer:19, farmer:20, renzo:20:conv_ppi,
  renzo:20:ppi_csm, woosley:21, farag:22}, illustrating the
theoretical consensus existing across authors, stellar evolution codes
(KEPLER and MESA), and a variety of assumptions.

\section{Observational predictions for (P)PISN}\label{sec:obs_predictions}

Despite being well understood from a theoretical perspective,
unambiguous observational confirmation of PISN is still lacking, and
the situation is even more unclear for P-PISN. This is likely due to a
combination of (\emph{i}) rarity of sufficiently massive and metal
poor progenitors in the local universe (see \Secref{sec:mass_ranges})
and (\emph{ii}) variety of predictions overlapping with observed
phenomenology from other transients (e.g., \citealt{woosley:17,
  woosley:19, renzo:20:ppi_csm, woosley:21}).

\subsection{What to look for?}\label{sec:pulse_types}

\begin{figure}[htbp]
  \centering
  \includegraphics[width=\textwidth]{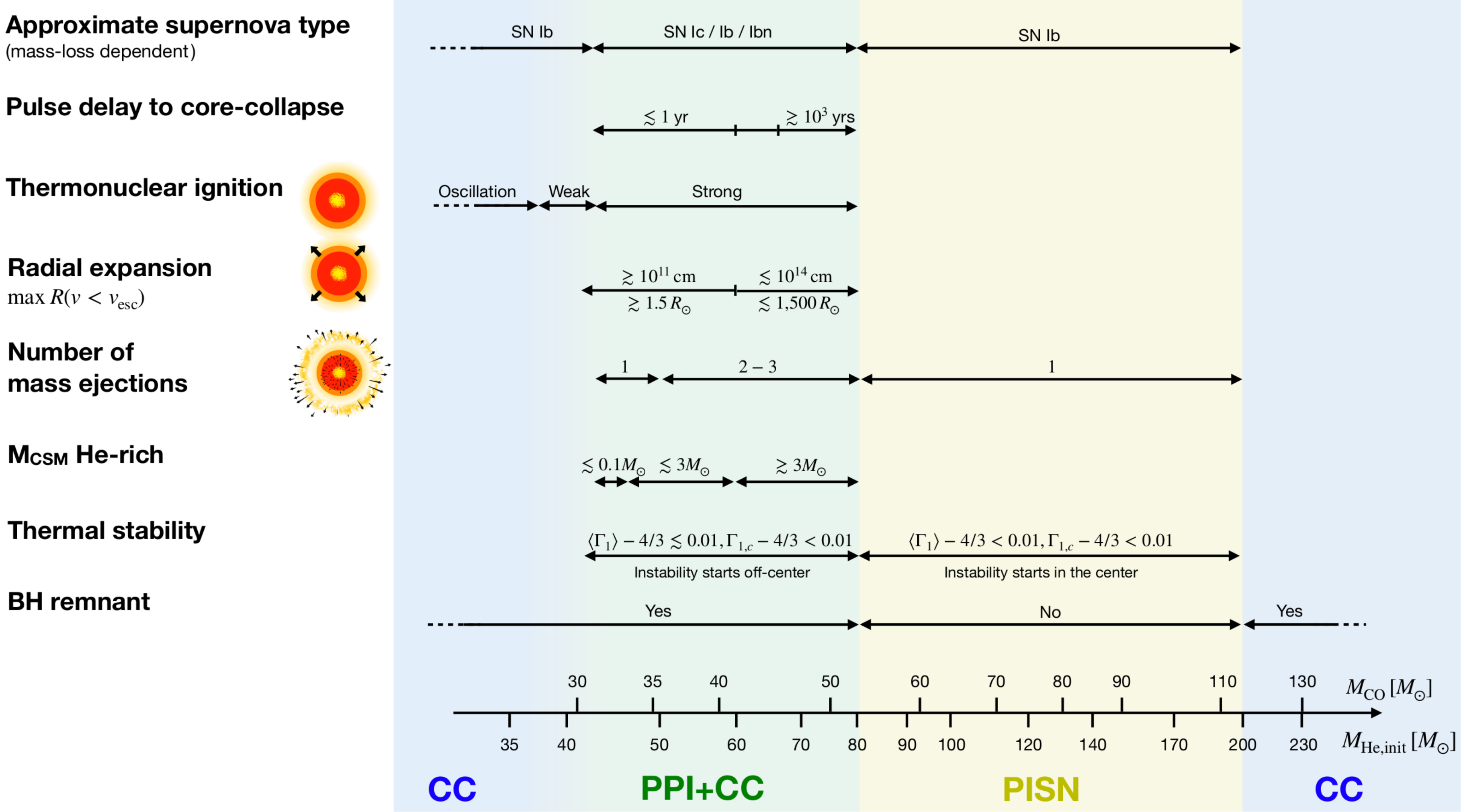}
  \caption{Overview of (P)PISN predicted phenomenology from
    \cite{renzo:20:ppi_csm}. This overview is based on bare He cores,
    the background colors match those in
    \Figref{fig:PISN_BH_mass_gap}. $M_\mathrm{CSM}$ indicates only the
    He-rich circumstellar material produced, since
    \cite{renzo:20:ppi_csm} did not include H-envelopes. The entire
    mass of the H-rich envelope should be added to those values in
    case it was retained until the first pulse (see also
    \Secref{sec:fate_envelope}). $v_\mathrm{esc}$ indicates the local
    escape velocity.}
  \label{fig:obs_summary}
\end{figure}

Depending on the observable considered, one can define a P-PISN
``pulse'' in different ways, which impacts the range of masses that
experience (P)PISN \citep{renzo:20:ppi_csm}. \Figref{fig:obs_summary}
shows an overview of the predicted phenomenology across a wide range
of initial He core masses from \cite{renzo:20:ppi_csm},discussed in
more detail below.

\subsubsection{Weak pulses and neutrino bursts}

At the lower mass end, pulses start as oscillations in the central
temperature and density \citep[e.g.,][]{woosley:17, paxton:18,
  woosley:19, marchant:19, renzo:20:ppi_csm}. These are symptoms of
\emph{local} instability that do not trigger a \emph{global} response
and do not impact the outer layers of the star. The only potential
observable of these ``weak pulses'' is through bursts in the neutrino
(mostly from cooling) emission, which increases at every maximum of
the density during the oscillations. However, because of the lack of
progenitors within the horizon for current and planned neutrino
detectors \citep[e.g.,][]{farag:20}, this is not a promising detection
strategy.

\subsubsection{Moderate pulses and radial expansion}

Increasing the core mass considered, pulses become more
energetic. Before they can lead to significant mass loss
($\gtrsim 1\,M_\odot$), there is a mass range where pulses cause
significant radial expansion, but the vast majority of the mass
remains bound to the star ($v$ remains lower than the escape velocity
$v_\mathrm{esc}$). This could be observed as large amplitude
photometric variations -- although (P)PISN is not the only
physical process that can induce those, and again the rarity of
progenitors in the local Universe is a limiting factor.

Potentially more promising is the detection of the signature of binary
interactions triggered by these large radius variations
\citep[e.g.,][]{marchant:19}. For example, if the companion is a
compact object, large radius variations caused by P-PISN pulses could
drive a high mass-transfer rate onto these, leading to copious X-ray
emission. It could also induce dynamically unstable mass-transfer,
although the larger the radial expansion, the lower the density of the
expanded material, and the less likely these interaction would have
dramatic orbital effects \citep{marchant:19}.

\subsubsection{Strong pulses and mass loss}

Moving further up in CO core mass, P-PISN start removing large amount
of mass, and eventually full-disruption by PISN removes \emph{all} the
mass, leaving no compact remnant behind. This is the regime that has
received most attention historically \citep[e.g.,][]{barkat:67,
  fraley:68}. The precise amount of mass loss and the shape of the
$M_\mathrm{BH}$-to-$M_\mathrm{CO}$ map are still actively being
studied \citep[e.g.,][]{woosley:17, renzo:20:conv_ppi,
  renzo:20:ppi_csm, mehta:22, farag:22, renzo:22, hendriks:23} and is known to be
sensitive to nuclear reaction rates \citep[e.g.,][see also
\Secref{sec:unknown_nuc}]{takahashi:18, farmer:19, farmer:20,
  mehta:22, farag:22} and the numerical resolution
\citep[e.g.,][]{farmer:19, farag:22}.

Overall, more massive cores experience more violent instabilities,
producing stronger thermonuclear explosions and more mass loss
\citep[][]{woosley:07, woosley:17, marchant:19, farmer:19, farmer:20, renzo:20:ppi_csm,
  renzo:22, farag:22}: within the mass-range for strong pulses (see
\Secref{sec:mass_ranges}) the lowest mass BHs come from the initially
more massive cores which experience more mass loss. However, there is
a minimum BH mass that can be reached. \cite{marchant:19} calculated
that cores capable of losing sufficient mass to drop their total mass
below $\sim{}10\,M_\odot$ would also produce sufficient amount of
$^{56}\mathrm{Ni}$ that the remaining $\sim10\,M_\odot$ of mass would
be unbound by the decay of this isotope\footnote{We note however that
  \cite{marchant:19} models employed a small nuclear reaction network,
  known to introduce large uncertainties in the $^{56}\mathrm{Ni}$
  yields, see also \cite{farmer:16, farmer:19, renzo:20:ppi_csm}.}.

\begin{figure}[htbp]
  \centering
  \includegraphics[width=0.5\textwidth]{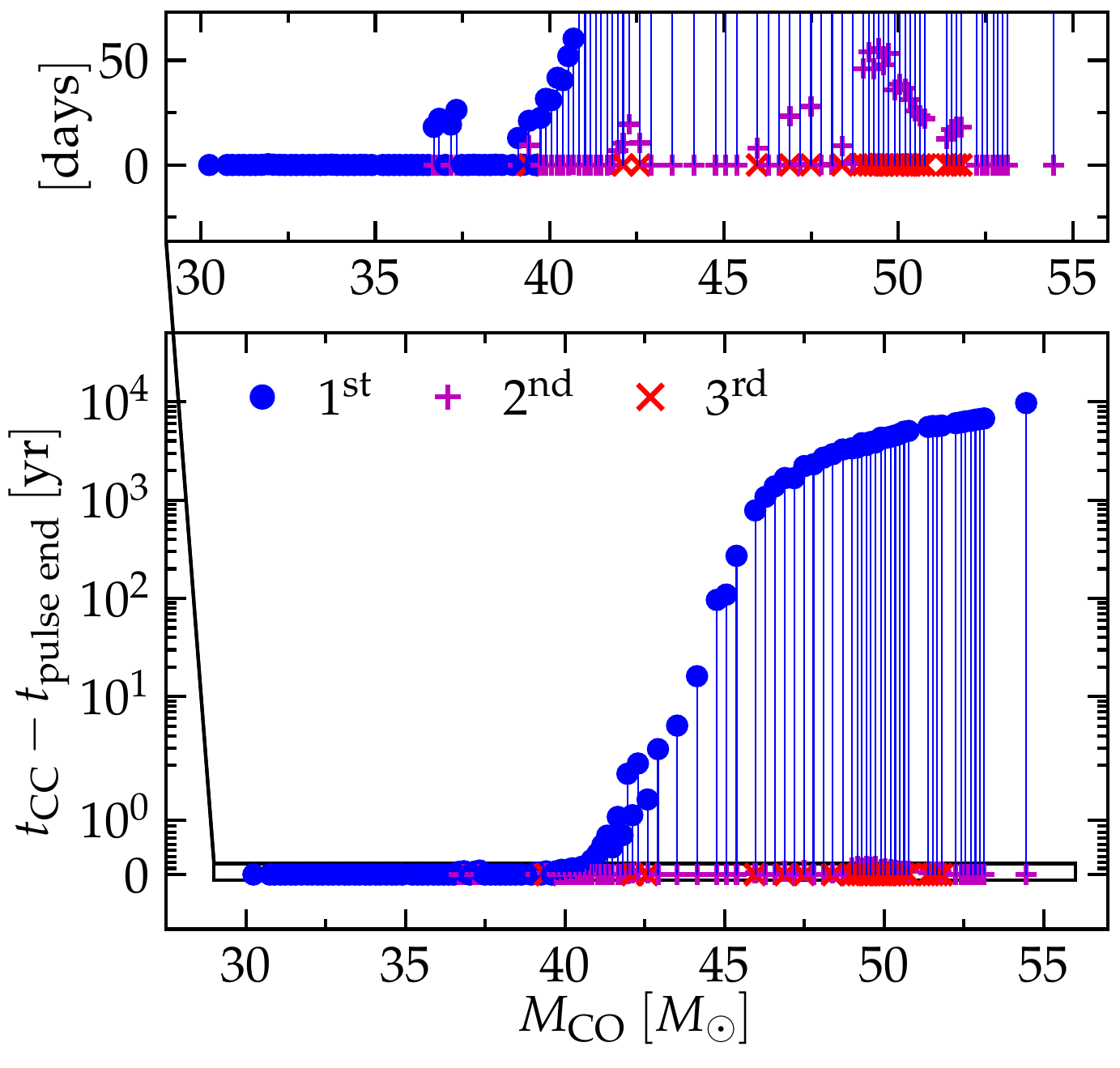}
  \caption{Time delay between dynamically distinct mass ejection
    events from \cite{renzo:20:ppi_csm}. At the low-mass end, many
    stars result in delay times of months, potentially creating
    (He-rich) circumstellar material detectable with flash
    spectroscopy.}
  \label{fig:time_delays}
\end{figure}

To determine the number of mass-ejection events in strong P-PISN, that
is, how many times a star goes through the loop in
\Figref{fig:cartoon} stellar evolution codes need to carefully define
the mass ejection events. \cite{renzo:20:ppi_csm} and \cite{umeda:20}
both found with completely independent computational approaches that
the number of distinct mass-ejection events (step 4b in
\Figref{fig:cartoon}) is $\lesssim 3$. The time-delay in between
pulses, determined by the timescale for the post-pulse relaxation
(step 5 in \Figref{fig:cartoon}) is a strong function of the CO core
mass, as \Figref{fig:time_delays} shows. More massive cores that
experience stronger pulses are also driven farther from equilibrium
and take a longer time to relax. At the most massive end, the delay
time between core-collapse and the end of a mass-ejecting pulses
$t_\mathrm{CC}-t_\mathrm{pulse\ end}$ reaches $\sim10^5\,\mathrm{yr}$,
that is roughly the Kelvin-Helmholtz timescale of stars of this mass
accounting only for their photon luminosity. This is because the pulse
expanded their core to the point where neutrino cooling emission shuts
off. Conversely, at the low mass end, many interpulse time of weeks to
months can be found: this can lead to bright transients if subsequent
pulses collide with each other \citep[e.g.,][]{woosley:07, woosley:17,
  renzo:20:ppi_csm} or if the final core-collapse (CC) event produces
an outflow.

Within each strong pulse, the ejecta have some velocity structure, but
overall the velocity is comparable to the escape velocity from the
surface ($v_\mathrm{esc}\simeq 10-100\,\mathrm{km\ s^{-1}}$ for red
supergiants and $\simeq1000\,\mathrm{km\ s^{-1}}$ for blue and/or
stripped progenitors). This can produce faster ejecta in the second
and third pulse (especially if the first pulse occurred in a expanded
red supergiant, while the second and third occur from a now stripped
and much more compact star), leading to collisions and electromagnetic
signals powered by conversion of kinetic energy into radiation
\citep[e.g.,][]{woosley:07, woosley:17, woosley:19, renzo:20:ppi_csm}.

\subsection{Electromagnetic transients}\label{sec:EM_obs}

\subsubsection{Pulsational pair instability supernovae: collision of
circumstellar shells}
%  P-PISNe may appear very different,
% arising from a series of failed explosions, where nuclear burning
% pulses ejected some of the outer envelope, but do not unbind the whole
% star, leading to a repeating series of pulses until the remaining star
% collapses into BH. The energy of each pulse, the mass ejected, the
% time between pulses, and the total number of pulses can vary greatly,
% leading to a wide diversity in light curves \citep{woosley:17,
% renzo:20:ppi_csm}.
% Moreover,

As discussed in \Secref{sec:pulse_types}, strong pulses can produce a
wide range of mass loss amount and timing, and power transients by
converting kinetic energy of the ejecta into radiation. This causes
the wide range of predictions for (P)PISN transients and consequently
the difficulty in finding non-controversial observational candidates.
Some are brief events (essentially a series of closely spaced pulses
that merge together) with irregular light curves, and some have
repeating events spread over many years.

Matter ejected in a pulse can have a velocity structure
\citep[e.g.,][]{woosley:17, renzo:20:ppi_csm}, leading to internal
collisions, but most importantly, the $2^\mathrm{nd}$ and
$3^\mathrm{rd}$ dynamically distinct ejection events can come from
deeper in the potential well of the star, that is with higher
$v\simeq v_\mathrm{esc}$ compared to the first ejection. This allows
them to catch up with the first pulse, producing a non-terminal event
of comparable luminosity as a terminal SN explosion by converting
kinetic energy of the gas into radiation. % P-PISNe are not powered by
                                          % radioactive decay;
                                          % instead, they
% are essentially circumstellar interaction powered SNe, arising when
% fast ejecta from a pulse overtake slower ejecta from the previous
% pulsation. The luminosity arises when the ensuing shock heats the
% material, converting bulk kinetic energy onto heat and light.
In this sense, P-PISN have some overlap with SNe of Types IIn and Ibn
\citep{smith:17}, but with different expected velocities seen in
spectra. This may even produce super-luminous events, but this is not
necessary: many P-PISN don't reach absolute magnitudes below -21
\citep{woosley:17, kozyreva:17}. \cite{woosley:07} and \cite{woosley:17}
calculated with KEPLER light curves for P-PISN showcasing the wide
phenomenology that theory predicts. Nevertheless, specific trends in
mass are expected, as exemplified in \Figref{fig:time_delays} for the
time delay between the pulses and final collapse, and may become
testable in the future \citep{renzo:20:ppi_csm}.

We emphasize that if the final collapse produces some, even weak,
outflow \citep[e.g.,][]{kuroda:18, burrows:23, powell:21, rahman:22},
the final blast wave can also potentially interact with previously
ejected shells, leading after ``SN impostor'' events to a final
terminal transient \citep[e.g.,][]{woosley:22}.

In another chapter in this Encyclopedia, \cite{dessart:24} reviewed
interaction-powered SNe, of which P-PISN may represent an extreme case
(in terms of progenitors and mass in the circumstellar shells).

\subsubsection{Pair instability supernovae: radioactive power}

The situation is less varied for PISNe, expected to be singular events
that obliterate the star in only one pulse (step 4a in
\Figref{fig:cartoon}), leaving no compact remnant. These are
thermonuclear SN powered by radioactive decay from a large synthesized
mass of $^{56}$Ni buried inside a large mass of stellar ejecta. The
% full disruption is possible because of the thermonuclear explosion
% that will produce large masses of
mass of $^{56}\mathrm{Ni}$ synthesized can exceed by orders of
magnitude the production of this isotope in core-collapse SNe. This
result in very high luminosity transients (up to
$\sim{}10^{44}\,\mathrm{erg\ s^{-1}}$, \citealt{kasen:11}). However,
the range of $^{56}\mathrm{Ni}$ masses spans $\sim 0.05-60\,M_\odot$
\citep[e.g.,][]{woosley:07, farmer:19, renzo:20:ppi_csm} so not all
PISN are necessarily super-luminous SNe \citep[e.g.,][]{kozyreva:17}.
In fact, the more massive the progenitor CO core, the higher the
synthesized $^{56}\mathrm{Ni}$ mass \citep[e.g.,][]{woosley:17,
  renzo:20:ppi_csm}, so assuming the CO core mass scales with the
initial total mass, the initial mass function (IMF) disfavors
super-luminous PISNe.

H-rich PISN are expected to have long-lasting light curves determined
by long radiation diffusion timescale across the ejecta
\citep{kasen:11, dessart:13, kozyreva:17}. H-poor transient instead do
not necessarily have large opaque ejecta masses, and fast and bright
transient may be produced \citep{kozyreva:17}. In both cases, the
light curves are expected to be smooth (no prior eruptions pollute the
circumstellar environment).

\section{Observed transients}\label{sec:obs_EM}

Several transient surveys are searching the sky for SNe and other
types of time variable phenomena, and it is obviously interesting to
consider whether transients triggered by the pair instability are
among the observed population of transient sources being detected in
these surveys. The intrinsic rate of pair-instability transients
should be rare compared to normal SNe, % . As noted above, the pair
% instability is only relevant in stars with very high initial mass,
% which are rare
because of the slope of the
IMF. % Traditionally, PISNe were expected to be limited to
% environments with very low metallicity \citep[e.g.,][]{langer:07}, due
% to the expectation that metallicity-dependent line-driven stellar
% winds would be strong enough to significantly reduce a star's mass
% during its main sequence lifetime, and therefore reduce its He core
% mass, allowing even the most massive stars to avoid the pair
% instability in local environments.
However, % for Pop III stars and other massive stars
at extremely low metallicity, the IMF may be skewed to higher masses
because of inefficient cooling during star formation
\citep[e.g.,][]{zinnecker:07}, and stellar winds are weaker, possibly
making (P)PISN more common. Thus, it has long been hoped that {\it
  JWST} might be able to see signs of PISNe from Pop III stars at very
high redshift \citep[e.g.,][]{whalen:13, regos:20}.

On the other hand, recent decades have seen significant reductions in
empirical estimates of line-driven wind mass-loss rates for hot stars
\citep{smith:14}. % good point.... and also for red supergiant wind mass-loss rates \citep{beasor:20,beasor:21} \mr{not sure that matters: by the time  stars are RSG their He core mass is set, it's the MS mass loss that  matters most to prevent P-PISN}.
In addition, untargeted transient
surveys have found many examples of unusual luminous transients, often
in very faint (presumably low-metallicity) host galaxies, and
observers have sometimes appealed to pair instability events for the
unusual observed explosions. In this section we review some of the
transients for which the pair instability has been invoked as a
potential explanation.

From an observational perspective, it is useful to discuss true PISNe
and P-PISNe separately, as the expected transients are very different
beasts, despite arising from the same trigger mechanism
\citep[e.g.,][]{woosley:17}. Below we discuss a number of events for
which the PI was suggested as a potential cause, and we comment on how
well they agree with expectations for PISNe and P-PISNe. Lists of
candidate events have also been compiled in \cite{renzo:20:ppi_csm}
(focused on H-poor events) and \cite{hendriks:23}.

\subsection{PISN candidates}\label{sec:PISN_candidates}

SN 2006gy:~This was the first recognized super-luminous" supernova
(SLSN). In addition to a very high peak luminosity, it had a slowly
rising, smooth light curve, and a PISN was suggested as one possible
explanation for its origin \citep{smith:07}; although interaction with
a massive shell of circumstellar material (CSM) produced by a massive
star eruption was considered more likely \citep{sm:07,smith:07}.
SN~2006gy was a Type IIn event, indicating that it had narrow H
emission lines from H-rich CSM. In its late-time decline, SN~2006gy
faded too quickly as compared to the sustained radioactive decay tail
that would have been expected from the large mass of $^{56}$Ni
powering the main peak in the PISN hypothesis
\citep{smith:08,miller:10}. This disfavored the PISN hypothesis,
although the P-PISN was a suggestion for the origin of the strong CSM
interaction (see \Secref{sec:P-PISN_candidates}).

SLSNe Ic: Around the same time SN~2006gy was being interpreted,
another class of SLSNe emerged, which had H- and He-poor spectra
(hence, classified as Type Ic-like spectra) and lacked the narrow
lines from CSM \citep{quimby:11}, with SN~2005ap being the prototype
\citep{quimby:07}. The situation with PISNe and this class is very
reminiscent of SN~2006gy above: although the PISN hypothesis was
initially discussed because of their high peak luminosity and smooth
lightcurves, they fade too quickly at late times to be explained by
radioactive decay from a large mass of $^{56}$Ni, and thus, true PISN
was rejected as an explanation \citep{quimby:11}. The currently
favored explanation is a H-free core-collapse SN that has added power
from magnetar spin-down \citep{woosley:10,kb:10}, and CSM interaction
via a P-PISN (see \Secref{sec:P-PISN_candidates}) may also be viable.

SN2007bi: This was a H-poor SLSN, similar in some respects to the
SLSNe~Ic discussed above. Unlike those objects, however, it faded at a
slower rate that was consistent with radioactive decay from a large
mass of $^{56}$Ni, and was proposed to be a true PISN
\citep{galyam:09}. However, spectral synthesis models predicted that
true PISNe will be quite red and cool because of strong line
blanketing from the large synthesized mass of Fe-group elements,
whereas SN~2007bi was very blue, disfavoring the PISN hypothesis
\citep{dessart:13,moriya:19}. These models also predicted more slowly
evolving light curves than SN~2007bi. Instead, CSM interaction appears
to be a viable explanation for SN~2007bi \citep{moriya:10,moriya:19}.

SN2018ibb:~Recently, \cite{shulze:24} described long
($\gtrsim$1000\,days) photometric and spectroscopic folloup of this
H-poor SLSN, and proposed it as a very strong PISN candidate based on
its slowly evolving light curve and spectrum, months-long rise time,
relatively slow ejecta velocities (from Fe\,II lines), peak bolometric
luminosity $\gtrsim10^{44}\,\mathrm{erg \ s^{-1}}$ corresponding to a
radiated energy $E_\mathrm{rad}\gtrsim10^{51}\,\mathrm{erg}$. The
transient has sufficiently faded to detect the host dwarf galaxy with
metallicity $\sim Z_{\odot}/4$. These observations are all in
agreement with theoretical prediction for PISN from very massive stars
producing tens of solar masses of $^{56}\,\mathrm{Ni}$, making it one
of the best candidates to date. However, this transient also showed
cobalt lines -- unexpected because of line blocking in PISN models,
and evidence for CSM interaction and a blue excess. The presence of
CSM is not naturally explained by PISN models, although it should be
noted that the theoretically unexpected presence of CSM occurs in many
non-exotic SNe as well (cf.~\cite{dessart:24} chapter of this
encyclopedia).

\subsection{P-PISN candidates}\label{sec:P-PISN_candidates}

As discussed above, P-PISN should arise from a range of initial masses
that is lower than the mass range giving rise to true PISNe, and thus,
we might expect P-PISNe to be somewhat more likely among the observed
transient population. On the other hand, the possible observed
outcomes of multiple shell collisions from P-PISNe are more diverse
than for PISNe \citep{woosley:17}, perhaps making them harder to
differentiate from other CSM interaction transients. Below, we mention
some examples discussed in the literature.

SN 2006gy: In addition to being the first proposed example of a true
PISN, SN~2006gy was also suggested as a possible example of a P-PISN
\citep{woosley:07,smith:10}, due to its high luminosity, slow
evolution, and observational signatures of strong CSM interaction. Its
massive inferred CSM shell ($\sim$20 $M_{\odot}$;
\citealt{sm:07,woosley:07}) ejected only a decade before discovery
\citep{smith:10} cannot be explained by any stellar wind scenario, but
could have arisen from a previous pair pulsation that ejected much of
the H envelope. The mass and energy budgets, as well as the
approximate appearance of the lightcurve (with some caveats), are
plausibly accounted for with a P-PISN model \citep{woosley:07}.
SN~2006gy was a Type IIn event, meaning that it had narrow lines from
slow H-rich CSM. In most cases, the first pulse is the most energetic
\citep{woosley:17, renzo:20:ppi_csm}, typically removing the remaining
H envelope (if any is left, see also \Secref{sec:fate_envelope}) in an
explosive blast with bulk expansion speeds of a few thousand km
s$^{-1}$ \citep{woosley:17, renzo:20:ppi_csm, woosley:22}. However, in
some models of P-PISNe for SN~2006gy with long delay times between
pulses, slower CSM speeds of $\sim$200 km s$^{-1}$ are seen initially,
with the observed pre-shock velocities rising as the SN fades. This is
in quite good agreement with spectra of SN~2006gy, which showed
pre-shock CSM expansion speeds of a few hundred km s$^{-1}$, rising
with time as the shock expanded outward as if running through a
self-similar %Hubble
flow rather than a steady wind \citep{smith:10}. Overall, the P-PISN
mechanism appears to give a good explanation for SN~2006gy, although a
predicted very bright precursor event was not detected, and other
mechanisms for pre-SN mass loss are not ruled out.

SN2009ip: This was a Type IIn supernova observed in 2012
\citep{mauerhan:13}, but before its final SN event it was seen as a
very luminous and likely very massive progenitor that had a series of
non-terminal LBV-like precursor outbursts \citep{smith:10ip}. Based on
its precursor outbursts and strong CSM interaction in the spectrum,
plus an integrated radiated energy of only $\sim$10$^{50}$ erg, a
P-PISN was one proposed scenario for this object
\citep{fraser:13,pastorello:13}. However, significant asymmetry
inferred from polarization and other observed properties indicated
instead that the 2012 event was a 10$^{51}$ erg exposion that was
typical for a core-collapse SN from a blue supergiant
\citep{mauerhan:14,smith:14ip}. Moreover, the broad-line spectrum of
the underlying SN ejecta (i.e. ``underneath" the narrow H lines) was
nearly identical to that of SN~1987A at many phases, showing broad H P
Cygni profiles that indicated a large mass of H-rich ejecta
\citep{smith:14ip}. As noted above for SNe IIn in general, retaining a
massive H envelope even after a series of major pulses is inconsistent
with a P-PISN for this object.

iPTF14hls: This object was dubbed ``The Impossible Supernova''
\citep{arcavi:17hls}, due to its high-luminosity multi-peaked light
curve that persisted for over 600 d while its spectrum hardly changed
at all. It also had a detected precursor eruption in the 1950s. A
P-PISN would potentially account for the mass budget and multiple
eruptions \citep{vignagomez:19, wang:22}, but as \citet{arcavi:17hls} noted,
the presence of H lines in the ejecta were problematic for a P-PISN,
similar to the cases of SN~2009ip and normal SNe IIn as discussed
below. Later spectra revealed narrow emission and double-peaked
intermediate-width emission, which were interpreted as a toroidal CSM
interaction region that was overtaken by the ejecta, powering the
sustained luminosity from within \citep{andrews:18}. These authors
noted that pre-SN binary interaction \citep{sa:14} would be a
plausible explanation for the origin of the asymmetric CSM, rather
than a P-PISN.

SN 1961V: This has always been considered a very enigmatic explosion.
Together, $\eta$ Car and SN~1961V comprised \citet{zwicky:64}'s
oddball class of ``Type V'' supernovae, although it would likely have
been classified as Type IIn had that designation existed at the time
\citep{smith:11}. SN~1961V had a very irregularly shaped, slowly
evolving light curve and somewhat narrow emission lines suggesting
expansion speeds of 2000-3700 km s$^{-1}$
\citep{bertola:63,zwicky:64,bg:71}. It also had a very luminous
progenitor star that was detected decades before the SN, which resided
in a giant H~{\sc ii} region and was estimated to have a very high
initial mass of $M_{\rm ZAMS} \approx 240 M_{\odot}$
\citep{goodrich:89}, although this initial mass may be lower because
of revised extinction, distance estimates, and/or the presence of
undetected companions. Although its vague
similarity to the eruption of $\eta$ Car led it to be considered as an
LBV or supernova impostor \citep{goodrich:89,filippenko:95,hds:99},
its peak luminosity was much brighter than any other SN impostor and
seems to have more in common with SNe IIn \citep{smith:11}. Comparing
the observed progenitor, historical lightcurve, late-time flux, and
reported velocities to model expectations, \citet{woosley:22} found
that P-PISN events from 100-115 $M_{\odot}$ progenitors give an
excellent match to all available data. Importantly, P-PISN models that
best matched the light curve also had progenitor stars with the
approximately the correct luminosity. Altogether, \cite{woosley:22}
concluded that SN~1961V was very likely a P-PISN, and is difficult to
explain with any other explosion mechanism.

Other SLSNe IIn and SNe IIn: In general, SNe IIn and especially SLSNe
IIn require extreme pre-SN mass loss that is difficult to explain with
steady stellar winds \citep[see][for a review of interacting
SNe]{smith:17}. As a result, the P-PISN is often cited as a possible
mechanism to create the dense CSM. While SLSNe~IIn are admittedly
quite rare, regular SNe~IIn are far too common (roughly 8-9 \% of all
core-collapse SNe; \citealt{smith:11}) to all be caused by the P-PISN,
which are limited to the most massive stars, and should be $<$1\% of
the core-collapse SN rate. While it is difficult to rule out any
individual event on these grounds alone, we can be confident, based on
rates, that the vast majority of SNe~IIn do not arise from P-PISNe.
Most SNe IIn and SLSNe IIn have the difficulty that their CSM is
consistently too slow, with CSM expansion speeds typically around
80-200 km s$^{-1}$, and generally not showing
increases in speed indicative of a self-similar flow in the CSM.
Moreover, in many cases, spectra of SNe IIn and SLSNe IIn show broad H
emission lines from the unshocked SN ejecta during the decline from
peak (refs). This is at odds with the P-PISN mechanism, since the
first pulse is expected to be the most energetic and should easily
remove the remaining H envelope. This means that the subsequent fast
SN ejecta that are overtaking any CSM should be H-free in a P-PISN.
This probably rules out the P-PISN mechanism for most SNe IIn and
SLSNe IIn \citep{woosley:22,smith:24}.

Transients compatible with P-PISN predictions have also been claimed
among the H-free SNe, too many to provide a complete list here. Below
we highlight some of the most notable candidates.

SN 2010mb: This was a H- and He-free transient (type Ic) showing large
ejecta masses and a long decay inconsistent with radioactive decay
\citep[e.g.,][]{ben-ami:14}. Spectroscopic signatures indicated the
presence of a dense and slow moving CSM, suggesting that the
progenitor had shed mass from the CO core -- making P-PISN the leading
model to explain this transient \citep{ben-ami:14}.

SN 2006jc: This was a Type Ibn supernova that was interacting with
He-rich/H-poor CSM. Although it was not an extraordinarily luminous
event, a P-PISN was mentioned as a possible origin for the CSM based
on the fact that it had a detected precursor outburst 2 yr before the
SN was discovered \citep{pastorello:07,yoshida:16}. If this were a
P-PISN, the star must have lost most of its H previously in its
evolution. The mass and energy budget would have been on the low end
for P-PISN models, implying a He core mass of 30-40 M$_{\odot}$
\citep{woosley:17}, although models in that range also predict a large
number of weak rapid pulses, rather than only 2 pulses separated by 2
yr (i.e., longer dealys between pulses of years correlate with larger
ejecta masses and larger pulse energies than observed estimates for
SN~2006jc). Nevertheless, the sensitivity of these predictions to stellar evolution
uncertainties (see \Secref{sec:known_unknowns}) makes a P-PISN
interpretation still possible. % For this object, a P-PISN may therefore be less likely.

iPTF16eh: \cite{lunnan:18} studied this SLSN and detected through
light echoes the presence of a O-rich CSM shell compatible with the
ejection of core-material by a previous P-PISN pulse.

SN 2016iet: \cite{gomez:19} interpreted this event as the final
explosion of a CO core embedded in H-poor CSM produced by previous
P-PISN pulses. While the timing inferred for the pulse ($\sim$ decade
prior to the SN detection) fits, the amount of H-poor CSM mass
required for the lightcurve is challenging to reconcile with existing
P-PISN models and the total mass of the expected progenitors.

\section{Indirect evidence}\label{sec:indirect}

Direct, unambiguous detection of transients interpreded as (P)PISN
have been challenging, thus many authors have focused on indirect
searches based on the signatures (P)PISN may leave in stars formed
from their ejecta and/or integrated signal of distant sources.
% Given the theoretical robustness of the (P)PISN mechanism (see
% \Secref{sec:known_unknowns}), a
Any indirect evidence for very massive stars in the early universe or
at very low metallicity (top heavy initial mass function -- e.g.,
\citealt{schneider:18}, super-massive BH formation from direct
collapse of supermassive stars -- e.g., \citealt{greene:20}, diffuse
X-ray background -- e.g., \citealt{sartorio:23}, etc.) can also be
used to support indirectly the occurrence of these explosions, but
definitive conclusions are still lacking.

One clear predicted signature is in the nucleosynthesis of these
explosions. For P-PISNe most of the ejecta will come from the H-rich
envelope (if any, see \Secref{sec:fate_envelope}), the He core, and
possibly outer-layers of the CO core -- except for (poorly understood,
cf.~\Secref{sec:unknown_conv}) mixing during the pulse propagation,
these are unlikely to eject much newly synthesized material.
Conversely, PISNe return all of their material to the host galaxy ISM,
and a larger fraction of the ejecta will be nuclear processed material
synthesized in the explosion \citep[e.g.,][]{woosley:02, takahashi:16,
  takahashi:18, farmer:19}. Since these explosions occur in a hot, and
low density CO core without free neutrons, this is expected to produce
a specific pattern in the nuclear yields: neutron-rich isotopes with
odd number of nucleons (which would require more neutrons) are highly
disfavored, resulting in a characteristic ``odd-even'' pattern in the
yields \citep[e.g.,][]{truran:70, woosley:02, takahashi:18}. We
emphasize that while the composition and temperature of the
thermonuclear explosions in white-dwarfs powering SNIa are similar to
those in (P)PISN, the density is much higher in the former leading to
electron captures and formation of neutron-rich isotopes that prevent
a strong odd-even effect in thermonuclear explosions of white dwarfs
\citep[e.g.,][and references therein]{gronow:21}. % , leading to
% more efficient photodisintegrations and changes in the composition of
% the ejecta. \N{This, in turn, leads to a weaker odd-even effect for
%   SNe Ia than is expected for PISNe (right?)}.

The characteristic lack of ``odd elements'' with a neutron excess may
be detectable in the composition of the next-generation stars formed
from the gas enriched by PISN. So far, unambiguous detections are
lacking. Traditional searches have focused on the spectroscopic
analysis of very low metallicity stars in the galactic halo, and have
produced several claimed detections \citep[e.g.,][]{aoki:14, xing:23}.
However, these remain controversial and potentially explainable
invoking only core-collapse SN yields \citep[e.g.,][]{takahashi:18,
  thibodeaux:24}. In general, only nuclear yields from KEPLER models
\cite{woosley:02, woosley:07, woosley:17, woosley:19} are available
for these analyses, which makes them vulnerable to unquantified
systematic uncertainties. Moreover, the focus on metal-poor low mass
stars may actually cut out candidate stars, if between the PISN
explosions and the formation of descendant stars, CCSNe can also
pollute the interstellar material, diluting the nucleosynthetic
signature of PISN.

\section{The gravitational-wave window}\label{sec:GW_obs}

\begin{figure}[hbp]
  \centering
  \includegraphics[width=0.5\textwidth]{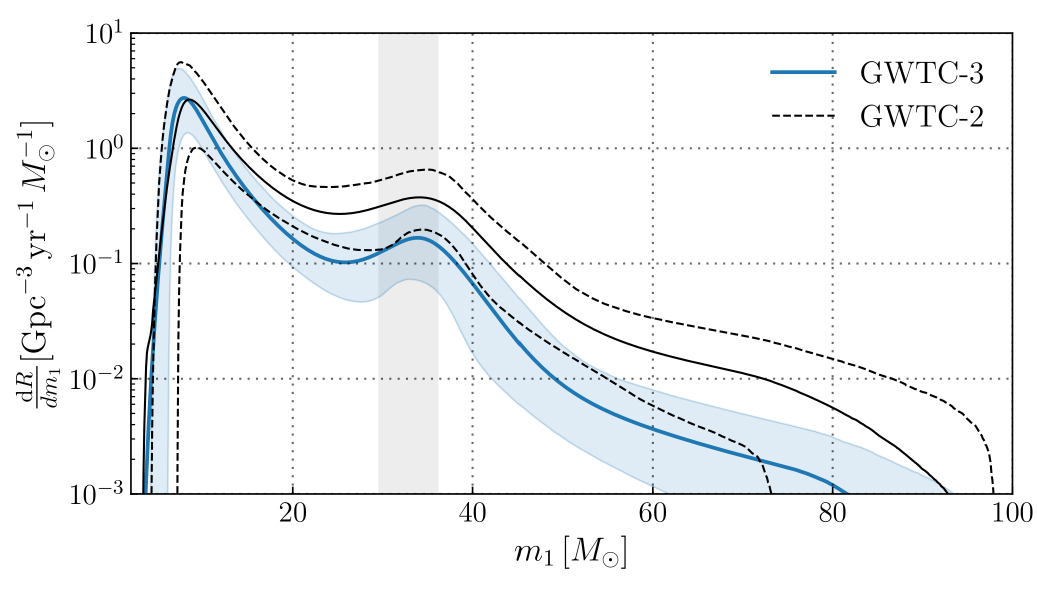}
  \caption{Binary BH merger rate inferred from GW merger detected as
    of the publication of the third GW source catalog in
    \cite{GWTC3_pop}, as a function of the mass of the most massive BH
    in the binary ($m_1$). The curves shown are the results of the
    ``powerlaw+gaussian peak'' model fit to the population. Black
    lines show the results from the second GW catalog \citep{GWTC2}, while
    blue lines show the most recent results to date \citep{GWTC3}.}
  \label{fig:BBH_mass_dist}
\end{figure}

The direct detection of GW from the coalescence and merger of binary
BHs opened a new way to study the distribution of remnants from
massive star evolution. Because (P)PISN can decrease the mass of BHs
resulting from the evolution of stars (due to mass-loss during the
pulses, step 4b in \Figref{fig:cartoon}) and/or prevent BH formation
completely (in the case of full PISN, step 4a in
\Figref{fig:cartoon}), the direct detection of GW has revived the
interest in these explosions \citep[e.g.,][see also
\Secref{sec:BH_mass_gap} and references therein]{woosley:17, spera:17,
  talbot:18, marchant:19, stevenson:19} and the potential signature
they may imprint on the BH mass distribution, which could also be a
``standardizable siren'' for cosmological applications
\citep[e.g.,][]{farr:19}.

However, less than a decade after the first direct detection of a GW
merger \citep{first_detection} and with $\sim{}100$ binary BH detected
to date \citep{GWTC3}, a clear signature of (P)PISN in GWs is also
lacking. \Figref{fig:BBH_mass_dist} shows the bias-corrected binary BH
merger rate inferred in \cite{GWTC3_pop} as a function of the mass of
the most massive BH in the merging binary $m_1$.
\Figref{fig:BBH_mass_dist} shows a tail extending to $\sim90\,M_\odot$
and a feature for BH masses of $\sim{}35\,M_\odot$. While this figure
shows a parametric (powerlaw + gaussian) model where such a feature is
forced to be a peak, that a feature around $\sim{}35\,M_\odot$ exist
is robustly found with different parametric models
\citep[e.g.,][]{GWTC2_pop, edelman:21} and non-parametric analyses
\citep[e.g.,][]{ray:23, callister:24}.

While several studies predicted a pile-up of BHs at the bottom edge of
the theorized PISN BH mass gap because of mass loss from P-PISN
\citep[e.g.,][]{woosley:17, spera:17, talbot:18, stevenson:19,
  marchant:19, farmer:19, farmer:20, renzo:20:ppi_csm, renzo:22,
  farag:22}, $\sim{}35\,M_\odot$ is much lower than predictions from
stellar evolution models \citep[e.g.,][]{woosley:17, woosley:19,
  marchant:19, farmer:19, farmer:20, renzo:20:ppi_csm, mehta:22,
  farag:22, hendriks:23}.

In \Secref{sec:known_unknowns} we review the known uncertainties that
could affect stellar evolution predictions, and although these may
allow for moving the predicted lower edge of the PISN BH mass gap as
low as 35\,$M_\odot$, this cannot be done without introducing some
tension elsewhere. For example, increasing the rate of certain nuclear
reaction rates (namely, \cag, see \Secref{sec:unknown_nuc} and
\citealt{takahashi:18, farmer:20, costa:21, mehta:22, farag:22,
  hendriks:23, croon:23:rates, golomb:23}) can put the lower edge of
the PISN BH mass gap at $\sim\,40\,M_\odot$ (see
\Figref{fig:mehta_gap}, and \citealt{farmer:20, mehta:22, farag:22}).
However, this would require a variation of more than $3\sigma$ in this
rate, in tension with recent nuclear laboratory measurements of this
reaction \citep[e.g.,][]{deboer:17, shen:23}. Moreover, shifting to
lower (CO core) masses the PISN BH mass gap causes an increase in the
predicted rate of (P)PISN (because of the increased probability of
forming sufficiently massive cores according to a stellar initial mass
function), and thus increasing the tension with the scarsity or lack
of transients robustly identified as (P)PISN
\citep[e.g.,][]{hendriks:23}. For these reasons, several studies have
disfavored the interpretation of the feature at $\sim{}35\,M_\odot$ as
a signature of (P)PISN \citep[e.g.,][]{hendriks:23, golomb:23,
  briel:23} and alternative explanation are currently highly debated.

\Figref{fig:BBH_mass_dist} also shows a tail of BH masses extending
all the way to $\sim90\,M_\odot$. We discuss in
\Secref{sec:known_unknowns} (and specifically in
\Secref{sec:fate_envelope}) systematic uncertainties in stellar models
that may lead to such BH masses, although again it is hard to
reconcile those without creating some tension elsewhere. Moreover,
several individual GW events showed at least one BH in the predicted
gap (if not both, e.g., for GW190521, \citealt{GW190521}~although see
also \citealt{fishbach:20}) seem to be at odds with the predictions on
(P)PISN. As we discuss in \Secref{sec:unknown_bin}, accretion onto the
first-born BH cannot explain these events: it would need to be
highly super-Eddington, and most importantly, it would result in
significant orbital widening, preventing GW-driven inspiral and merger
within the age of the Universe \citep{vanson:20}.
These very massive BHs could be reconciled if the PISN BH mass gap
moves towards higher (CO core) masses -- helping also with the lack of
electromagnetic detections of (P)PISN -- which could happen because of
several reasons from changes in the input nuclear reaction rates
\citep[e.g.,][and \Secref{sec:unknown_nuc}]{takahashi:18, farmer:20,
  farag:22, mehta:22}, treatment of convection and boundary mixing
\citep[e.g.,][and \Secref{sec:unknown_conv}]{farmer:19,
  renzo:20:conv_ppi, umeda:20}, or the fate of the H-rich envelope for
models just below the threshold for (P)PISN \citep[e.g.,][see also
\Secref{sec:fate_envelope}]{vink:21, sabhahit:23, winch:24}. Another
potential explanation proposed by \cite{shibata:21} is that the signal
from these events may not be due to a binary BH merger in the first
place (although see \citealt{siegel:22}). Mass-loss at BH formation
post P-PISN, usually not accounted for in stellar evolution
simulations \citep{renzo:22}, could remove mass in BHs forming above
the gap (step 4c in \Figref{fig:cartoon}) polluting the gap ``from
above'' \citep{siegel:22}. Finally, there is the possibility that
these events come from dynamical channels and involve $2^\mathrm{nd}$
(or higher) generation BHs which are not direct products of stellar
evolution \citep[e.g.,][]{romero-shaw:20}.

\Figref{fig:BBH_mass_dist} also shows how the inferred merger
rate as a function of BH mass is still changing (from the second to
third catalog published), and the mass estimates for each individual
event are sensitive to the priors assumed in the analysis
(\citealt{fishbach:20}). One should expect further surprises as
ground-based GW observations continue and space-based GW observation
start in the next decade.

% \N{(something i don't understand about the predictions for this gap --- be definition, all these merging BHs evolved in binaries.  even if there is some limit on the BH masses from formation, can't any of these be smeared out as a BH accretes mass from its companion star during its evolution as a HMXB before that second star collapses to a BH?  has someone included this effect?  )}

\section{Known uncertainties}\label{sec:known_unknowns}

We now discuss the many known uncertainties in the evolution of
(P)PISN progenitors. \cite{farmer:19} performed an extensive survey of
stellar evolution uncertainties to assess their impact on the BH
masses resulting from (P)PISN. Despite how exotic these explosions may
seem, this revealed a surprising robustness of the results within the
framework of the evolution of bare He cores. In the following
subsections, we focus on the major theoretical uncertainties for
(P)PISN identified in \cite{farmer:19} and subsequent studies
\citep[][]{farmer:20, renzo:20:conv_ppi, woosley:21, mehta:22, farag:22,
  shen:23}. These add to uncertainties in the general evolution of
massive stars discussed elsewhere \citep[e.g.,][]{vink:15, farmer:16,
  renzo:17, davis:19, josiek:24}.

\subsection{Nuclear reaction rates}\label{sec:unknown_nuc}

(P)PISN are caused by a thermonuclear explosion in the star following
the collapse caused by pair-production decreasing the radiation
pressure support. Therefore, it may not be surprising that
uncertainties in the nuclear reaction rates impact (P)PISN. Most of
the energy release during the thermonuclear explosion is from
$^{16}\mathrm{O}+^{16}\mathrm{O}$ (when this isotope is present, e.g.,
\citealt{dessart:13, marchant:19}), and the biggest uncertainty does
not come from the explosive burning itself, but rather from the
reactions determining the amount of fuel available to an explosion --
in other words, the pre-instability reactions determining the C/O
ratio in the core. These are the reaction creating $^{12}\mathrm{C}$,
\citep[$3\alpha$,][]{hoyle:54,farag:22, luo:24} and the reaction
destroying it \citep[\cag,][]{kunz:02, deboer:17, shen:23}. The impact
of the former on P-PISN was studied in \cite{farag:22}, while the
impact of the latter is the focus of \cite{takahashi:18,
  farmer:19, farmer:20, costa:21, farag:22, mehta:22, kawashimo:23}.

Specifically, \cite{takahashi:18} proposed that a lower \cag rate
could lead to a sufficiently low $^{16}\mathrm{O}$ mass fraction in
the core, pushing the mass range for (P)PISN upward and explaining the
lack of observations of these transients. This result was
independently confirmed by \cite{farmer:19} and physically explained
in \cite{farmer:20}: with higher C/O ratio a thick convective C
burning shell appears in the models, protecting the core from
instability by making it effectively evolve as a lower mass CO core.
This effect was also seen in the models from \cite{costa:21} and
\cite{woosley:21}. % , making it robust across 4 different and independent
% stellar evolution codes.

\begin{figure}[htbp]
  \centering
  \includegraphics[width=0.5\textwidth]{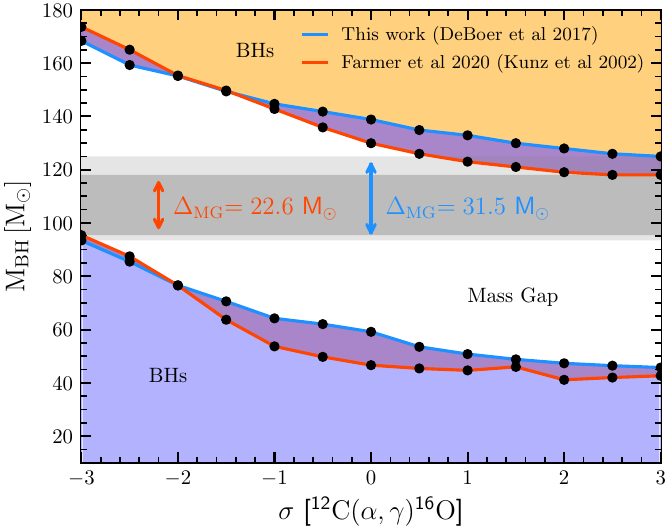}
  \caption{Lower and upper edge of the PISN BH mass gap as a function
    of the rate of \cag from \cite{mehta:22} (blue) and
    \cite{farmer:19} (red). The caption also indicates the nominal
    rate adopted (\citealt{deboer:17} in \citealt{mehta:22} and
    \citealt{kunz:02} in \citealt{farmer:19}, respectively), and the
    x-axis represents the number of standard deviations from the
    nominal rate. The gray horizontal band represent BH masses that
    cannot form from single bare He cores, regardless of the \cag
    rate.}
  \label{fig:mehta_gap}
\end{figure}

However, \cite{mehta:22} realized that publicly available tables for
the rate of \cag as a function of temperature were under-resolved:
interpolation errors in stellar codes were larger than the $3\sigma$
experimental uncertainties (see Figure~7 in \citealt{mehta:22}). Their
study % in collaboration with nuclear physicist
ammended this and updated the results of \cite{farmer:20} and made
new, more dense, nuclear reaction rates tables available for the
community. \Figref{fig:mehta_gap} shows the upper and lower edge of
the theoretical PISN BH mass gap from \cite{mehta:22} as a function of
the adopted \cag rate.

\cite{farag:22} further built upon these results, showing the
degeneracy between the assumed nuclear reaction rates and the temporal
and spatial resolution required to obtain numerically
converged\footnote{By numerical convergence we mean results not
  dependent on the choice of spatial and temporal discretization
  \citep[e.g.,][]{farmer:16, farag:22}.} results. Furthermore, they
extented the investigation of nuclear physics uncertainties to the
$3\alpha$ reaction that dominates the production of carbon.

\cite{kawashimo:23} also showed, albeit with a small 21-isotope
nuclear reaction network, that the total amount of radioactive
$^{56}\mathrm{Ni}$ produced in PISN full disruptions and the upper
boundary of the theorized PISN BH mass gap (see
\Secref{sec:BH_mass_gap}) are also sensitive to the C/O ratio and
thus the \cag rate, with lower rates producing less radioactive
material.

We emphasize that the rates (typically tabulated in stellar evolution
codes) are still debated in the nuclear physics community
\citep[e.g.][]{kunz:02, deboer:17, kibedi:20, shen:23, luo:24}, and from a
stellar perspective, any rate affecting the C/O ratio is expected to
impact the (P)PISN process.

\subsection{Treatment of convection}\label{sec:unknown_conv}
% \todo{convection \cite{renzo:20:conv_ppi}}

Because of its inherent multi-dimensional nature in relation to
turbulence, convection is always a major source of uncertainty in
stellar evolution calculations. Most models, including of (P)PISN
progenitors, rely on mixing length theory (\citealt{bohm-vitense:58,
  jermyn:22}), which is an ``effective theory'' describing the fully
developed convection in steady-state. While this is sufficient for
most stellar applications, for (P)PISN progenitors, an additional
complication arises: during the dynamical phase (steps 3 and 4 in
\Figref{fig:cartoon}) the evolutionary timescale % \N{in the core}
is shorter than the
convective turnover time % \N{in the envelope (?)} % no really
% globally! This is unlikely any stellar evolution phase (maybe He
% flash is similar)
and the assumption of steady-state breaks down
\citep[e.g.,][]{renzo:20:conv_ppi}. A self-consistent
spherically-symmetric treatment of the convective acceleration,
describing how turbulent energy transport turns on/off in a dynamical
and stratified medium is needed to treat the \emph{local} energy transport
throughout the structure as the pulse propagates.

% \N{(question - i guess i'm confused here. why woud convection matter
% *during* a pulse? a pulse is a shock moving through the star and
% ejecting material from the surface on a dynamical timescale. or do
% you mean during the time in between pulses, during the time when the
% core relaxes and resumes burning? or what? it is hard to see how
% convection can be significant in bringing energy to the surface to
% be radiated away more than a shock that ejects mass suddenly... but
% in the longer time between puklses is sorta makes sense to me.)}
% It's more of a technical problem in a sense: the question is how
% energy is transported *during* the pulse propagation. That takes
% only a dynamical timescale (but it's usually 1000s of timesteps in
% the code): that is too fast for us to use MLT. If you still naively
% use it, like we did originally, we have that convection turns on/off
% instantaneously and in the end you get stronger pulses at the onset
% of P-PISN. This is most likely a numerical artifact. If you use a
% model for how convection turns on/off based on simple dimensional
% analysis, you see it turns on and stays on, and leaks energy,
% weakening the pulse. A better model is needed because we don't know
% how to treat macroscopic turbulent energy flow in a dynamical medium
% -- that's not what MLT is for, but that's what you have during pulses.

If convection efficiently develops during the propagation of a (P)PISN
pulse, it can carry energy away until it is radiatively leaked at the
surface, weakening the mass loss driven by the pulse, and influence
the final BH mass \citep{renzo:20:conv_ppi}. This is particularly
relevant for the lower-mass end of the (P)PISN regime, where different
simple treatment of the convective acceleration based on
\cite{unno:67, arnett:69, gough:77} compared to instantaneous
development of convection result in significantly different amount of
mass loss. This happens around predicted BH masses of
$\sim35\,M_\odot$, a regime now probed directly by both GW detections
\citep{LIGOpop, hendriks:23} and astrometry \citep{gaiaBH3}, which calls for more
detailed studies of the development of convection during (P)PISN
pulses.

A further uncertainty related to convection is convective boundary
mixing \citep[e.g.,][]{umeda:20, costa:21, vink:21}. This impacts
directly the initial mass range and thus predicted rate of (P)PISN:
more mixing during the main-sequence leads to more massive He cores
and thus more (P)PISN for a given initial mass distribution of stars.
Conversely, a smaller amount of mixing will results in smaller cores
per a given initial total mass, which can prevent (P)PISN directly,
but also influence the stellar structure \citep{umeda:20}, the C/O
ratio in the core (see \Secref{sec:unknown_nuc}), the appearance of
the star and thus its wind mass loss rate \citep{vink:21,
  sabhahit:23}. The description of convective boundary mixing remains
an open problem and while one-dimensional algorithms for convective
boundary mixing informed by three-dimensional simulations exist
\citep[e.g.,][]{anders:22, johnston:24}, they are not designed for the
very high mass regime relevant to (P)PISN.

\subsection{Impact of rotation}\label{sec:unknown_rot}

\begin{figure}[htbp]
  \centering
  \includegraphics[width=0.5\textwidth]{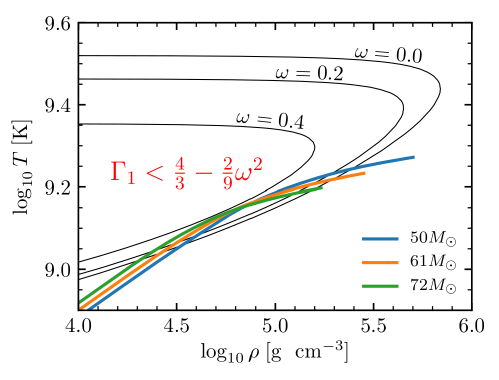}
  \caption{Instability region on the $T-\rho$ plane modified by the
    contribution of centrifugal forces to hydrostatic equilibrium from
    \cite{marchant:20}. $\omega$ indicates the rotational frequency.
    Colored tracks are non-rotating ($\omega=0$) stellar profiles from
    \cite{marchant:19} labeled by their total initial He core mass
    (\citealt{marchant:19} did not include any H-rich envelope in their
    calculations).}
  \label{fig:gamma_rot}
\end{figure}

Rotation, much like convection, breaks the spherical symmetry, and for
this reason it is a source of uncertainty in one-dimensional stellar
calculations. In the context of (P)PISN, one can expect three main
effects of rotation:

\begin{enumerate}
\item Mixing during the hydrostatic evolution changing the
  relation between core masses and total initial mass
  \citep[e.g.,][]{maeder:00, chatzopoulos:12, chatzopoulos:13,
    umeda:24}. Similar to increasing main sequence convective boundary
  mixing, high rotation with efficient associated mixing can increase
  the expected rate of occurrence of cores sufficiently massive to
  experience (P)PISN.
\item Centrifugal force changing dynamical equilibrium considerations
  (see \Figref{fig:gamma_rot}). Overall, based on spherically-averaged
  stellar evolution models, this produces a rather small effect on the
  final BH masses. The maximum BH mass below the PISN BH mass gap
  shifts upward by $\sim{}4\%$ only when accounting for angular
  momentum transport \citep{spruit:02} in the star during the
  pre-pulse hydrostatic evolution \citep{marchant:20}.
\item Dynamical effect of rotation \emph{during} the pulse generation
  and propagation (steps 3 and 4 in \Figref{fig:cartoon}). This has
  not yet been thoroughly investigated (although see
  \citealt{glatzel:85}).
\end{enumerate}

\subsection{Development of asymmetries}
\label{sec:unknown_3D_effects}

Similar to the uncertainties discussed in \Secref{sec:unknown_conv}
and \Secref{sec:unknown_rot}, any other process breaking spherical
symmetry that cannot be captured by one-dimensional stellar evolution
codes is a source of uncertainty. Asymmetries may develop during the
thermonuclear ignition \citep{renzo:20:ppi_csm} and during the
propagation of a pulse \citep{chen:14, chen:23}.

The former has not
yet been thoroughly investigated, but it was suggested as a
possibility at the lower mass end of P-PISN in \cite{renzo:20:ppi_csm}
because of the competition between neutrino cooling and energy release
by the thermonuclear explosion. In fact, initially in the lowest mass
P-PISN models (that is, the ones with the highest core densities at
ignition), neutrino cooling dominates over the energy release by the
explosion. This means the region of the star where burning is
dominating is an off-center shell -- a situation reminescent of other
stellar thermonuclear explosions in white dwarfs which could lead to
seed asymmetries in the explosion that may be amplified during the
pulse propagation. % This has not yet been investigated in detail to
% determine observational consequences (if any). % feels like a repetition
% \N{(...would these actually do anything (i mean asymmetric ejection)
% as the shock goes out from the small O core to the surface?)}
% completely unexplored: we don't know!

The other possibility is the development of asymmetries due to
Rayleigh-Taylor instabilities during the pulse propagation. Starting
with a spherically symmetric explosion, \cite{chen:14, chen:23} showed
that these do occur, making multi-dimensional radiation hydrodynamics
simulations necessary to obtain detailed light-curves of (P)PISN, but
remain overall small in amplitude.

\subsection{Binary interactions}\label{sec:unknown_bin}

Most massive stars are born in binary systems \citep[e.g.,][]{sana:12,
  offner:23}, and the binary fraction does not depend on metallicity
\citep[e.g.,][]{moe:17} or maybe increases as metallicity decreases
(e.g., \citealt{price-whelan:20} based on Galactic low mass red
giants). Although the binary status of the most massive stars known to
date is unclear, these are found in a dense stellar cluster (R136 in
the 30\,Doradus region, \citealt{dekoter:97, crowther:10, crowther:16,
  brands:22}) where dynamical processes are likely to pair them into
binaries even if they formed as single \citep[e.g.,][]{fujii:11}. This
makes the study of binary interactions among (P)PISN progenitors
particularly important and presently lacking, especially in the
context of the impact of these explosions on BH masses and GW
astronomy.

So far, only calculations based on bare single He cores -- assumed to
mimic stars stripped of their envelope by a companion long before the
(P)PISN onset -- have been used to consider the impact of binary
interactions \citep[e.g.,][]{marchant:19, woosley:19, woosley:21}.
While this is a useful starting point to discuss, for example,
interaction of the expanded post-pulse star with hypotetical
companions which could power copious electromagnetic emission
\citep[][]{marchant:19},
it cannot account for the modifications of the core structure due to
the timing and amount of mass loss \citep[e.g.,][]{laplace:21}, or
address the impact of binary interactions on accretor stars
\citep[e.g.,][]{renzo:21, miszuda:21, renzo:23, wagg:24} or mergers
\citep[e.g.,][]{renzo:20:merger, costa:22, ballone:23}.

Nevertheless, rapid binary population synthesis calculations use
single star models \citep[e.g., from][]{woosley:19} or single bare He
core models \citep[e.g., from][]{marchant:19, farmer:19} to calculate
BH masses \citep{spera:19, stevenson:19, renzo:22, hendriks:23}. Since these models
come from numerical simulations distinct from the ones informing BH
mass estimates for core-collapse SNe
\citep[e.g.,][]{fryer:12,fryer:22}, this can introduce numerical
artifacts resulting in overproduction of BHs at certain masses
\citep[e.g.,][]{vanson:22}. To avoid this issue, models fitting for
the amount of mass lost in detailed stellar evolution simulations of
(P)PISN, rather than the resulting BH mass, should be preferred
\citep[e.g.,][]{renzo:22}.

Besides modifying the core structure with as-of-yet poorly explored
impact on (P)PISN, binary interactions can also alter the BH mass
function after the BHs form, through accretion. \cite{vanson:20}
showed that this is not expected to produce a detectable effect in the
GW mass distribution of BHs: if sufficient accretion can occur,
binaries widen too much to become viable GW sources; conversely if
accretion is limited by radiation-pressure in the accretion flow
(Eddington-limited accretion), the amount of mass accreted is small
compared to existing uncertainties in this BH mass regime.

\subsection{Fate of the H-rich envelope (if any)}\label{sec:fate_envelope}

Very massive stars have a few ways to potentially lose their H-rich
envelope long before developing a CO core, either through steady
stellar winds \citep[e.g.,][]{lucy:70, castor:75, smith:14, vink:15,
  renzo:17, beasor:21, vink:21, sabhahit:23, decin:24}, binary
interactions \citep[e.g.,][]{langer:12, sana:12, gotberg:17,
  gotberg:18}, and/or eruptive mass loss
\citep[e.g.,][]{smithowocki:06,smithLBV:11,smith:14,
  jiang:15,jiang:18, cheng:24}. Moreover, the main-sequence
core-to-envelope mass ratio increases with stellar mass, leading to
quasi-homogeneous evolution at finite metallicity
\citep[e.g.,][]{yusof:13, kozyreva:17}. These arguments are used by
many authors to study (P)PISN from bare He cores, since the
pair-instability is likely to often start after the H is lost
\citep[e.g.,][]{heger:02, woosley:19}, and it is particularly true
when focusing on GW sources from isolated binary evolution
\citep[e.g.,][]{marchant:19, farmer:19, farmer:20, woosley:21,
  farag:22}, which require final separations of the binary
incompatible with an extended H-rich envelope -- although in dense
stellar environments, BHs may form separately from single star
evolution and be dynamically paired afterwards
\citep[e.g.,][]{dicarlo:19, kremer:20}.

Nevertheless, should a star retain its H-rich envelope until the onset
of the pulses, it is then guaranteed to lose it in the first pulse.
This can be understood with an energetic argument: typically, extended
envelopes have binding energies of $\sim0.6-1 \times 10^{50}$, while
pulses release $\gtrsim 10^{50}-10^{51}\,\mathrm{erg}$ sufficient to
unbind the H-layers during the first pulse
\citep[e.g.,][]{woosley:07}, as shown by calculations from
\cite{woosley:07, kasen:11, woosley:17, leung:19, renzo:20:ppi_csm}.
In case the entire envelope is not ejected, it will be bloated
\citep[e.g.,][]{marchant:19, renzo:20:ppi_csm}, lowering its chances
to survive the subsequent pulses. Even for cases where a stellar merger
adds back a H-rich envelope late in the progenitor evolution, it is
unlikely to survive the P-PISN \citep[e.g.][]{vignagomez:19}.

% \N{(comment - one other thing we could mention. it might also be
% possile for a PISN progenitor to GAIN a H enveope late in lifew if
% it ndergoes a late-phase merger. i don;t know how common this should
% be, but itmight be a possible way for a star to have a H envelope
% and be a PISN or p-PISN...still likely to be lost in first pulse, i
% guess.)} That is basically the scenario from Vigna-Gomez et al. 2019
% (cited elsewhere), they didn't calculate the pulses, but they said
% it should fly away
To determine the lower edge of the PISN BH mass gap
(\Secref{sec:BH_mass_gap}), the remaining issue is what happens to
stars just not quite massive enough to encounter the pair-instability
with an H-rich envelope. This could occur for single stars at
sufficiently low metallicity (preventing efficient wind mass loss,
\citealt{vink:21, sabhahit:23, winch:24}), or because of mergers in a
dynamical environment \citep[e.g.,][]{dicarlo:19, dicarlo:20a,
  dicarlo:20b, renzo:20:merger, kremer:20, costa:21, costa:22,
  ballone:23} creating a non-standard stellar structure with an
under-massive core and oversized envelope. If these envelopes are
extended, many mass loss mechanisms are possible at the final collapse
\citep[e.g.,][]{quataert:19, ivanov:21, antoni:22, burrows:23}.
If instead the envelope is sufficiently compact and manages to avoid
eruptive instabilities, and the core is sufficiently small to avoid
(P)PISN, it may be possible for it to collapse into a BH swallowing (a
large fraction of) the envelope \citep[e.g.,][]{renzo:20:merger}. One
possibility to keep the envelope sufficiently compact, blue, and thus
bound is to have small amount of core boundary mixing during the
H-burning main-sequence phase \citep[e.g.,][]{umeda:20}, possibly
coupled with smaller wind mass-loss rate \citep[e.g.,][]{vink:21}
leading to a small He core-to-envelope mass ratio that favors blue,
compact, stellar solutions \citep[e.g.,][]{arnett:89, langer:89}.
(P)PISN from these progenitor structure are briefly explored in
\cite{woosley:17} and still result in complete envelope removal at the
first pulse. PISN from blue supergiants are discussed in
\cite{kasen:11} and \cite{dessart:13}.

\subsection{Beyond-standard-model variations}\label{sec:beyond_SM}

The theory of (P)PISN explosions is rooted in particle physics
(cf.~\Secref{sec:intro_theory}), therefore, it is not surprising that
modifications to the standard model may have an impact on the
predictions for (P)PISNe. Several beyond-standard-model effects have
been included in stellar evolution simulations to test their impact on
the minimum CO core mass required for instabilities and on the
resulting BH masses. From a stellar evolution point of view, these
effects can be studied modifying either the energy generation,
hydrostatic equilibrium equations, and/or the equation of state.
Beyond-standard-model effects studied so far include:

\begin{itemize}
\item Dark matter annihilation \citep[e.g.,][]{ziegler:21, croon:23}.
  This adds an energy source to the star, whose spatial distribution
  depends on the assumed mass of the dark matter particle, and may
  prevent the instability, but only for very high dark matter
  densities \citep{croon:23}.
\item Addition of weakly interacting particles that can escape the
  star as soon as they are produced (e.g., ``hidden'' photons --
  \citealt{croon:20}). This adds an energy loss term throughout the
  evolution of the star, leading to faster nuclear evolution and
  larger C/O ratios in the core, producing effects similar to lowering
  the \cag rate.
\item Addition of more-strongly interacting particles (e.g., axions --
  \citealt{sakstein:22, mori:23}). If these couple with photons, this provides
  an extra channel to lose radiation pressure support on top of
  \Eqref{eq:pair_production}, exhacerbating the pair-production instability.
\end{itemize}

Whether the proposed
modifications are compatible with other existing constraints on
beyond-standard-model physics remains to be studied in detail.

\section{Conclusions}\label{sec:conclusions}

The study of (P)PISN started almost exactly 60 years ago
\citep{fowler:64}. Possibly, because of the lack of clear,
unambiguous, and uncontroversial observational confirmation, it is
characterized by surprising theoretical consensus in terms of the
physical evolution of single massive star cores. These are challenged
by recent GW detections and many solutions to this apparent
discrepancy have already been proposed and are presently hotly debated
while further gravitational and electromagnetic detections are
collected.

However, we live in a Universe where most massive stars are in
multiple systems \citep[e.g.,][]{sana:12, offner:23}, a fact seldom
included in (P)PISN progenitor modeling, and where the evolution stars we
observe with masses well below the threshold for pair-instability
still pose theoretical questions. These include their wind mass-loss rates
\citep[e.g.,][]{smith:14, vink:15, renzo:17, vink:21, sabhahit:23,
  winch:24}, their late evolutionary stages \citep[e.g.,][]{farmer:16,
  laplace:21, schneider:24, renzo:24} and final collapse
\citep[][]{powell:21, rahman:22}. The more massive the star, the more
significant theoretical uncertainties \citep[e.g.,][]{agrawal:22}, and
the more rare direct observations to improve our models.

Time-domain astronomy allows for probing farther, where progenitors
cannot be seen. Despite some initially exciting candidates for PISNe
from modern surveys, it remains unclear that we have actually observed
one. Very recently, \cite{shulze:24} put forward SN2018ibb as the best
candidate to date. For P-PISNe, there appears to be few very strong
case of a P-PISN (e.g.,~SN~2010mb, \citealt{ben-ami:14}, iPTF16eh,
\citealt{lunnan:18}, and SN~1961V, \citealt{woosley:22}), accompanied
by a few other candidates (SN~2006jc; SN~2006gy; and by extension,
some other SLSNe IIn if they do not show broad H lines). Transients
caused by the pair instability are therefore likely to be quite rare.
Whether this is consistent with expected rates for their initial mass
ranges remains somewhat uncertain, and may depend on the star
formation and evolution models. It may also depend strongly on our
ability to correctly diagnose a pair-instability event when we see it.
As noted above, true PISNe should be relatively straightforward to
identify, whereas the diverse outcomes of P-PISNe might likely be
misclassified and may still be lurking among samples of observed
transients.

\subsection{Future prospects}

The opportunities to shed light on the fate of the most massive stars
are plentiful. JWST is currently operating, so perhaps the lack of
unambiguous direct electromagnetic detections will change soon
\citep{hummel:12, whalen:13, regos:20}. Future missions such as ROMAN
\citep{moriya:23} and Euclid \citep{moriya:22, tanikawa:23} may also
find PISN and P-PISN transients. Ground based surveys such as
Rubin/LSST will also contribute, especially in the search for slowly
evolving transients such as those predicted for PISNe.

Moreover, GW detection of merging binary BH will continue. The
LIGO-Virgo-KAGRA observing runs O4 (ongoing at the time of writing)
and O5 (ending in 2029) will provide better statistics to understand
the population of \emph{merging} binary BHs (a very small fraction of
the total, but nevertheless informative). Farther in the future,
$3^\mathrm{rd}$ generation ground-based detectors promise to detect
every binary BH merger up to redshifts beyond the predicted formation
of the first stars -- possibly revealing a fuller picture.

Finally, the wide-spread availability of stellar evolution codes
capable of alternating between hydrostatic and dynamical phases of
evolution during pulses has allowed for the first time to start
exploring the theoretical and input physics uncertainties. The role of
pre-pulse binary interactions remain so far unexplored, but are likely
relevant for the formation of GW sources and observational transients,
and some relevant physics (mixing, nuclear reaction rates, opacities)
is still being updated, making this field ripe for advances.

\begin{ack}[Acknowledgments]%

  MR is grateful to D.~Hendriks, S.~Justham, L.~van~Son, R.~Farmer,
  and S.~E.~de~Mink, for many illuminating discussions on the
  (astro)physics of these explosions over the years, to P.~Marchant
  and especially R.~Farmer for invaluable help in making simulations
  of (P)PISN possible with a community-driven and open-source code
  (MESA), and to D.~Croon for feedback on beyond-standard-model
  modifications to (P)PISN. We also emphasize the pioneering role of
  S.~Woosley in developing the theory of (P)PISN which has been an
  inspiration throughout the years.

\end{ack}

\seealso{ \cite{marchant:19} provide at
  \url{https://zenodo.org/records/3786599} movies for the evolution of
  several quantities during the P-PISN of massive He cores. We also
  refer readers to \cite{lin:23, kuncarayakti:23, Aamer:24, sharma:24}
  for other transients discussed in the context of (P)PISN not covered
  above.}

\bibliographystyle{Harvard}
\bibliography{reference}

\end{document}